\documentclass[transmag]{IEEEtran}
\usepackage{latexsym}
\usepackage[dvipsnames]{xcolor}
\usepackage{graphicx}
\usepackage{amsfonts,amssymb,amsmath}
\usepackage{algorithmic}
\usepackage{hyperref}
\def\BibTeX{{\rm B\kern-.05em{\sc i\kern-.025em b}\kern-.08em T\kern-.1667em\lower.7ex\hbox{E}\kern-.125emX}}
\usepackage[utf8]{inputenc}
\usepackage[T1]{fontenc}
\usepackage{xcolor}
\usepackage{cite}
\usepackage{bm}
\DeclareMathOperator*{\maximize}{maximize}
\DeclareMathOperator*{\minimize}{minimize}
\ifCLASSOPTIONcompsoc
    \usepackage[caption=false, font=normalsize, labelfont=sf, textfont=sf]{subfig}
\else
\usepackage[caption=false, font=footnotesize]{subfig}
\fi
\begin{document}

\title{Energy Efficiency of Uplink Cell-Free Massive MIMO With Transmit Power Control in Measured Propagation Channel}
\author{Thomas Choi, Masaaki Ito, Issei Kanno, Jorge Gomez-Ponce,\\Colton Bullard, Takeo Ohseki, Kosuke Yamazaki, and Andreas F. Molisch
\thanks{%This paragraph of the first footnote will contain the date on which you submitted your paper for review. It will also contain support information, including sponsor and financial support acknowledgment. For example, ``This work was supported in part by the U.S. Department of Commerce under Grant BS123456''.
A part of this article was submitted to the IEEE SiPS 2021. The work of T. Choi, J. Gomez-Ponce, C. Bullard, and A. F. Molisch is supported by KDDI Research, Inc. and the National Science Foundation (ECCS-1731694 and ECCS-1923601). J. Gomez-Ponce is also supported by the Foreign Fulbright Ecuador SENESCYT Program.
}
%\thanks{The next few paragraphs should contain the authors' current affiliations, including current address and e-mail. For example, F. A. Author is with the National Institute of Standards and Technology, Boulder, CO 80305 USA (e-mail: author@ boulder.nist.gov).}
\thanks{T. Choi, J. Gomez-Ponce, C. Bullard, and A. F. Molisch are with the University of Southern California, Los Angeles, CA 90006 USA (e-mail: \{choit, gomezpon, ctbullar, molisch\}@usc.edu). J. Gomez-Ponce is also with ESPOL Polytechnic University, Escuela Superior Politécnica del Litoral, ESPOL, Facultad de Ingeniería en Electricidad y Computación, Km 30.5 vía Perimetral, P. O. Box 09-01-5863, Guayaquil, Ecuador.}
\thanks{M. Ito, I. Kanno, T. Ohseki, and K. Yamazaki are with KDDI Research, Inc, Saitama, Japan (e-mail: \{sk-itou, is-kanno, ohseki, ko-yamazaki\}@kddi-research.jp).}}

\IEEEtitleabstractindextext{\begin{abstract}%(The abstract should not exceed 250 words. It should briefly summarize the essence of the paper and address the following areas without using specific subsection titles.): Objective: Briefly state the problem or issue addressed, in language accessible to a general scientific audience. Technology or Method: Briefly summarize the technological innovation or method used to address the problem. Results: Provide a brief summary of the results and findings. Conclusions: Give brief concluding remarks on your outcomes. Clinical Impact: Comment on the translational aspect of the work presented in the paper and its potential clinical impact. Detailed discussion of these aspects should be provided in the main body of the paper.

%(Note that the organization of the body of the paper is at the authors' discretion; the only required sections are Introduction, Methods and Procedures, Results, Conclusion, and References. Acknowledgements and Appendices are encouraged but optional.)

Cell-free massive MIMO (CF-mMIMO) provides wireless connectivity for a large number of user equipments (UEs) using access points (APs) distributed across a wide area with high spectral efficiency (SE). The energy efficiency (EE) of the uplink is determined by (i) the transmit power control (TPC) algorithms, (ii) the numbers, configurations, and locations of the APs and the UEs, and (iii) the propagation channels between the APs and the UEs. This paper investigates all three aspects, based on extensive ($\sim$30,000 possible AP locations and 128 possible UE locations) channel measurement data at 3.5 GHz. We compare three different TPC algorithms, namely maximization of transmit power (max-power), maximization of minimum SE (max-min SE), and maximization of minimum EE (max-min EE) while guaranteeing a target SE. We also compare various antenna arrangements including fully-distributed and semi-distributed systems, where APs can be located on a regular grid or randomly, and the UEs can be placed in clusters or far apart. Overall, we show that the max-min EE TPC is highly effective in improving the uplink EE, especially when no UE within a set of served UEs is in a bad channel condition and when the BS antennas are fully-distributed.
\end{abstract}

\begin{IEEEkeywords}
Channel capacity, energy efficiency, massive MIMO, microwave propagation, power control, wide area measurements
\end{IEEEkeywords}
%Note: There should no nonstandard abbreviations, acknowledgments of support, references or footnotes in in the abstract.
}
\maketitle

\section{Introduction}
\subsection{Motivation}
Cell-free massive MIMO (CF-mMIMO), which combines various wireless communication system concepts such as mMIMO, ultra-dense networks, and cooperative multi-point (CoMP), exploits a large number of access points (APs) distributed across a wide area to reliably serve a large number of user equipments (UEs) while suppressing the inter-cell interference conventional cellular systems suffer from \cite{demir2020foundations}.
%They are anticipated to play a major role in 5G deployments, and the 3GPP NR standard provides many elements facilitating their use \cite{dahlman20215g}.
CF-mMIMO has significant performance advantages compared to traditional systems: the distributed nature of the antenna elements increases the reliability, which is a prerequisite for many Internet of Things (IoT) and mission-critical applications. Their distributed nature also makes them a natural fit for mobile edge computing. 

Important performance metrics in CF-mMIMO systems are the energy efficiency (EE) of the (battery-powered) UEs, which is mainly determined by the information transmission in the uplink phase, i.e., UEs to APs, as well as the spectral efficiency (SE) of this process. The use of transmit power control (TPC) can, to a certain degree, trade-off EE and SE: if interference from other UEs can be cancelled, increase of transmit power improves the SE of a selected UE, but decreases the EE, because SE increases logarithmically with transmit power, while energy consumption increases in an affine way. The situation is further complicated by the fact that different UEs experience different attenuation to the various APs, and interference between the signals of different UEs can impact the SE. 
%When the UEs transmit information to the APs during the uplink phase, the energy efficiency (EE) is an important performance metric, especially if the UEs are battery-powered and if the required uplink spectral efficiency (SE) is low. 
Hence, finding transmission algorithms, especially transmit power control (TPC) algorithms, that can maximize the EE at a target SE is very important, but nontrivial, and their performance assessment is challenging.

The performance of such algorithms is critically impacted by the propagation channels the system is operating in, as the wireless providers planning the deployment of CF-mMIMO systems require accurate and reliable statistics of the expected performance. Thus, it is important to test such TPC algorithms on realistic channel data. Such real-world data can only be obtained from extensive channel measurement campaigns. However, large measurement datasets for CF-mMIMO systems are scarce due to the complexity of setting up and operating a massive number of antennas simultaneously. To address this issue, we recently proposed that a large amount of channel data for CF-mMIMO systems can be measured using a compact channel sounder with a drone acting as a virtual array and released open-source channel data \cite{choi2021using}. 

\subsection{Related Works}
Within traditional mMIMO, various power allocation methods to optimize wireless system performance have been considered. In \cite{zhao2015analysis}, the trade-offs between the uplink SE and EE were analyzed through power models and simulations. In \cite{saatlou2018power}, a TPC scheme which optimizes both the SE and EE was developed. Furthermore, the ways to allocate power to both the data and pilots in order to maximize the SE was studied in \cite{saatlou2019joint} and an optimum number of base station (BS) antennas which can improve the downlink EE was studied in \cite{hao2018energy}. However, these works were focused on co-located massive MIMO system, where there is only one AP per cell.

There were also numerous CF-mMIMO studies which tackled the problem of improving the EE. In order to save energy at the APs, efforts were made to maximize the \emph{total downlink EE}, with various precoding methods and operating frequencies \cite{nguyen2017energy, ngo2018on, alonzo2018energy, tran2019first, jin2020spectral, qiu2020downlink}. Other works analyzed the downlink EE while maximizing the \emph{minimum downlink SE} per UE, in the cases of hardware impairments \cite{zhang2018performance}, in comparison to cellular systems \cite{yang2018energy}, or in relation to security \cite{alageli2020optimal}. For uplink, the EE was analyzed to maximize the \emph{minimum uplink SE} among the UEs \cite{bai2019max-min, zhao2020efficient, zhang2020analysis, demir2021joint, yan2021scalable}. There were also efforts to optimize the power coefficients for both the uplink and downlink jointly while seeking a balance between the EE and SE \cite{nguyen2020on, wang2020wirelessly, zhang2021spectral}.

In a recent conference paper \cite{ito2021}, we suggested the max-min EE method, which optimizes the power allocation to maximize the \emph{minimum uplink EE} over all UEs, at a given target SE. This algorithm improved EE for UEs with the lowest EE in comparison to the max-power and max-min SE algorithms. This is practically meaningful because all UEs want to have a sufficient lifetime on a battery charge. However, all the works mentioned above including our own \cite{ito2021} were based on simulated channel data achieved from statistical channel models. In \cite{wang2021live}, a CF-mMIMO testbed was developed between 16 APs and 16 UEs, but the environment was limited to indoors and the EE was not considered.
%While \cite{choi2021using} observed the uplink SE of CF-mMIMO systems using channel measurement data, it lacked analysis of EE and impact of TPC.

\subsection{Contributions}
To provide a more realistic assessment of TPC algorithms, and bridge the gap between the theory and practical implementation of CF-mMIMO, we apply three different TPC algorithms (max-power, max-min SE, and max-min EE) to a large number of {\em measured} propagation channel data at 3.5 GHz to analyze the trade-offs between the EE and SE for CF-mMIMO systems with varying numbers, configurations, and locations of the APs and UEs. The amount of data used for these analyses is very large, featuring $\sim$30,000 possible AP locations and 128 possible UE locations across a 200m$\times$200m area, providing statistical confidence of the evaluated performances in a realistic deployment setting. The current paper considers EE and SE for (i) various antenna arrangements including fully-distributed (single-antenna AP) and semi-distributed (multi-antenna AP) systems, (ii) AP distributions, where APs can be located on a regular grid or randomly, and (iii) UE distributions, i.e., the UEs can be placed in clusters, or far apart from each other. This is in contrast to the conference version of the current paper \cite{choi2021uplink}, where only single-antenna AP configuration and 8 UEs placed close to one another were considered for the analysis, using zero-forcing combining. We show that the max-min EE is very effective, especially when no UE within a set of served UEs is in a bad channel condition, minimum mean square error (MMSE) combining is applied, when more BS antennas are used in comparison to the number of UEs, when the BS antennas are fully-distributed evenly across the coverage area, and when the UEs are far apart in the case of distributed BS antennas.

\begin{figure}[!t]
\centering
\includegraphics[width=0.85\linewidth]{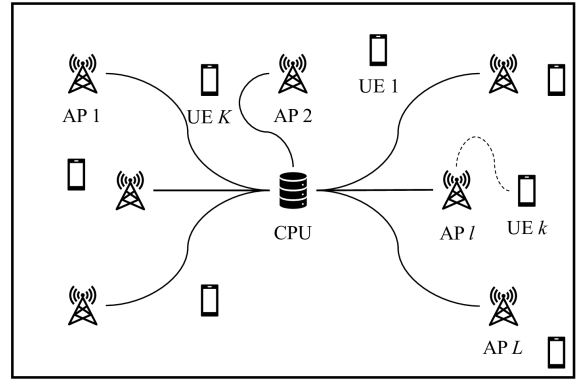}
\caption{A cell-free massive MIMO system: $K$ UEs are served by a BS composed of a central processing unit (CPU) and $M$ antennas. $L$ APs, each with $N$ antennas, are distributed across a coverage area ($M=LN$).}
\label{fig:cfmmimo}
\end{figure}

\section{System Model} \label{system}
We consider a CF-mMIMO system (\figurename~\ref{fig:cfmmimo}), where the BS is composed of $L$ APs with $N$ antennas each (the total number of BS antennas is hence $M=LN$). The APs are deployed in a selected area of service to support $K$ single-antenna UEs simultaneously in each time-frequency resource block.

\subsection{Channel Model}
The collection of complex channel coefficients between $M$ BS antennas and $K$ UEs is written into a matrix $\bm{\mathrm{H}}$. Indexing the BS antennas $m=1,...M$ and UEs $k=1,...K$, the channel coefficient $h_{m,k}=\bm{\mathrm{H}}(m,k)$ for frequency-flat fading is:

\begin{IEEEeqnarray}{rCl}
h_{m,k}&=&\sqrt{\beta_{m,k}}g_{m,k},%
\end{IEEEeqnarray}
where $\beta_{m,k}$ and $g_{m,k}$ are large- and small-scale fading of the corresponding links, respectively. If an AP has more than one antenna, we assume $\beta_{m1,k}=\beta_{m2,k}$,\footnote{Small-scale fading will be different for $m1$ and $m2$: $g_{m1,k}\neq g_{m2,k}.$} where $m1$ and $m2$ are two different antennas within AP $l$; in 3GPP parlance, the antennas are co-located in one AP.

\subsection{Uplink System Model}
 We consider the uplink case, when the UEs transmit signals to the BS. The received signal at BS antenna $m$ is:

\begin{IEEEeqnarray}{rCl}
y_m&=&\sqrt{\rho}\sum_{k=1}^K h_{m,k}\sqrt{q_k}s_k+z_m,%
\end{IEEEeqnarray}
where $s_k$ is a transmitted symbol of UE $k$ normalized to unit average power, $0 \leq q_k \leq 1$ is the transmit power coefficient, $z_m\sim\mathcal{N}_{\mathbb{C}}\mathopen{}\left(0,1\right)\mathclose{}$ is the normalized noise, and $\rho$ is the transmit SNR, i.e., the ratio of the maximum transmitted signal power to the noise power.\footnote{This transmit SNR must not be confused with SNR from the channel measurement in Sec. \ref{chan_meas} (the measurement only provides $\bm{\mathrm{H}}$). $\rho$ is obtained `manually', by 1) assigning arbitrary transmit power of the UE and 2) computing the noise power of the system with arbitrary bandwidth and noise figure.} Likewise, the collection of received signals at all $M$ BS antennas can be written as:

\begin{IEEEeqnarray}{rCl}
\bm{\mathrm{y}}=\sqrt{\rho}\bm{\mathrm{H}}\bm{\mathrm{Q}}^{1/2}\bm{\mathrm{s}}+\bm{z},%
\end{IEEEeqnarray}
where the dimension of the diagonal matrix $\bm{\mathrm{Q}}$ is $K\times K$ ($\bm{\mathrm{Q}}(k,k) = q_{k}$). Vectors $\bm{\mathrm{s}}$ and $\bm{\mathrm{z}}$ have the dimensions $K\times1$ and $M\times1$, respectively ($\bm{\mathrm{s}}(k)=s_k$ and $\bm{\mathrm{z}}(m)=z_m$). 

\subsection{Channel Estimation}
For channel estimation, $\tau^{\mathopen{}\left(\text{p}\right)\mathclose{}}$-length pilot resources from each UE are used within the coherence interval.\footnote{We assume all UEs use orthogonal pilot sequences in this paper.} Let $\sqrt{\tau^{\mathopen{}\left(\text{p}\right)\mathclose{}}}\bm{\mathrm{\varphi}}_k$ be the $\tau^{\mathopen{}\left(\text{p}\right)\mathclose{}}$-dimensional pilot sequence vector of UE $k$, where $\mathopen{}\left\|\bm{\mathrm{\varphi}}_k\right\|\mathclose{}^2=1$. Then, the received signal vector is written as:

\begin{IEEEeqnarray}{rCl}
\bm{\mathrm{y}}_m^{\mathopen{}\left(\text{p}\right)\mathclose{}}&=&\sqrt{\rho^{\mathopen{}\left(\text{p}\right)\mathclose{}}\tau^{\mathopen{}\left(\text{p}\right)\mathclose{}}}\sum_{k=1}^Kh_{m,k}\bm{\mathrm{\varphi}}_k+\bm{\mathrm{z}}_m^{\mathopen{}\left(\text{p}\right)\mathclose{}}.%
\end{IEEEeqnarray}

The MMSE channel estimate can then be written as \cite{ohurley2020comparison}:

\begin{IEEEeqnarray}{rCl}
\Hat{h}_{m,k}&=&\frac{\sqrt{\rho^{\mathopen{}\left(\text{p}\right)\mathclose{}}\tau^{\mathopen{}\left(\text{p}\right)\mathclose{}}}\beta_{m,k}}{\rho^{\mathopen{}\left(\text{p}\right)\mathclose{}}\tau^{\mathopen{}\left(\text{p}\right)\mathclose{}}\sum_{k'=1}^K\beta_{m,k'}\mathopen{}\left|\bm{\mathrm{\varphi}}_k^\text{H}\bm{\mathrm{\varphi}}_{k'}\right|\mathclose{}^2+1}\bm{\mathrm{\varphi}}_k^{\text{H}}\bm{\mathrm{y}}_m^{\mathopen{}\left(\text{p}\right)\mathclose{}}.%
\end{IEEEeqnarray}

\section{Performance Metrics}
To analyze the performances of different TPC algorithms in Sec. \ref{TPC}, we evaluate the SE and EE. We assume either maximum-ratio (MR) combining or MMSE combining on the BS side, where the weight matrices are expressed as:

\begin{IEEEeqnarray}{rCl}
\bm{\mathrm{w}}^{\mathrm{MR}}_k={\hat{\bm{\mathrm{h}}}_k},%
\end{IEEEeqnarray}

\begin{IEEEeqnarray}{rCl}
\bm{\mathrm{w}}_k^{\mathrm{MMSE}} =\rho q_k\bigg(\sum_{i=1}^K\rho q_i (\hat{\bm{\mathrm{h}}}_i\hat{\bm{\mathrm{h}}}_i^\mathrm{H}+\bm{\mathrm{C}}_i)+\bm{\mathrm{I}}_{LN}\bigg)^{-1}\hat{\bm{\mathrm{h}}}_k,%
% p_i (\hat{\bm{\mathrm{h}}}_i\hat{\bm{\mathrm{h}}}_i^\mathsmaller{\mathrm{H}}
\end{IEEEeqnarray}
where $\hat{\bm{\mathrm{h}}}_i$ is the estimate of channel vector for UE $i$, $\bm{\mathrm{h}}_i=\bm{\mathrm{H}}(:,i)$, and $\bm{\mathrm{C}}_i=\mathbb{E}\{\tilde{\bm{\mathrm{h}}}_i\tilde{\bm{\mathrm{h}}}_i^{\mathrm{H}}\}$ is the error correlation matrix, where the channel estimation error vector for UE $i$, $\tilde{\bm{\mathrm{h}}}_i$, is defined as $\tilde{\bm{\mathrm{h}}}_i=\bm{\mathrm{h}}_i-\hat{\bm{\mathrm{h}}}_i.$

The MR has the advantage over MMSE for its simplicity, as it can even be implemented locally per AP. The MMSE in contrast, has to be implemented centrally after collecting the channel data from all APs due to the matrix inversion process. Yet, the MMSE can cancel out the interference from other UEs, providing better performance at the cost of its complexity. It has been shown in the seminal paper of Marzetta \cite{marzetta2010noncooperative} that in the limit of very large arrays, the MR performance converges to that of MMSE; though experimental investigations with co-located arrays with 64 antennas have shown significant performance differences \cite{shepard2018argos}. As we will see in Sec. V, there also are significant performance differences between MR and MMSE in our (distributed) mMIMO arrays, even when the number of antennas is large. 
\subsection{Spectral Efficiency}
The SE of UE $k$ is:

\begin{IEEEeqnarray}{rCl}
S_k%&=&\log_2\mathopen{}\left(1+\text{SINR}_k\right)\mathclose{}\IEEEnonumber\\
\!=\!\log_2\!\mathopen{}\left(1\!+\!\frac{\rho q_k |\bm{\mathrm{w}}_k^\mathrm{H}\hat{\bm{\mathrm{h}}}_k|^2}{\rho\!\sum_{k'\neq k}^K\!q_{k'}\!|\bm{\mathrm{w}}_k^\text{H}\hat{\bm{\mathrm{h}}}_{k'}\!|^2\!+\!\bm{\mathrm{w}}_k^{\mathrm{H}}\bm{\mathrm{Z}}_k\bm{\mathrm{w}}\!+\!\mathopen{}\left\|\!\bm{\mathrm{w}}_k\!\right\|\mathclose{}^2}\right)\mathclose{}\!,\label{eq:SE}
\end{IEEEeqnarray}
where $\bm{\mathrm{Z}}_k=\rho\sum_{i=1}^{K}q_i\bm{\mathrm{C}}_i$.

\subsection{Energy Efficiency}
Based on~\cite{bashar2019energy}, the power consumption of UE $k$ is:

\begin{IEEEeqnarray}{rCl}
P_{k}&=&\Bar{P}q_k+P_{\text{U}},\label{eq:ee_rev}%
\end{IEEEeqnarray}
where $\Bar{P}$ is the maximum transmit power and $P_{\text{U}}$ is the required power to run circuit components at each UE. The EE of UE $k$ is then defined as:

\begin{IEEEeqnarray}{rCl}
E_k&=&\frac{\text{Bandwidth}\cdot S_k}{P_{k}}.\label{eq:ee_ue}%
\end{IEEEeqnarray}

\section{Transmit Power Control Algorithms} \label{TPC}
In this work, we consider three different types of uplink TPC algorithms: max-power, max-min SE, and max-min EE.

\subsection{Max-Power Method}
Max-power is the most simplistic method: each UE transmits with the maximum allowed power ($q_k=1$). It is not strictly a TPC method, but we use it as the baseline to be compared with other TPC algorithms. 

\subsection{Max-Min Spectral Efficiency Method}
Max-min SE is one of the most commonly used TPC methods in the CF-mMIMO literature, and aims to maximize the minimum SE among all UEs. The optimization problem is:

\begin{IEEEeqnarray}{ll}
\maximize_{\mathopen{}\left\{q_k\right\}\mathclose{}}&\min_{k=1,\dots,K}S_k\label{eq:prob}\\
\text{subject to }&0\le q_k\le1,k=1,\dots,K.\IEEEnonumber%
\end{IEEEeqnarray}

Since the SE is a logarithmic function increasing monotonically with the signal-to-interference-plus-noise ratio (SINR), the problem \eqref{eq:prob} can be reformulated as:

\begin{IEEEeqnarray}{ll}
\maximize_{\mathopen{}\left\{q_k\right\}\mathclose{},t}&t\label{eq:prob2}\\
\text{subject to }&t\le\text{SINR}_k,k=1,\dots,K\IEEEnonumber\\
&0\le q_k\le1,k=1,\dots,K.\IEEEnonumber%
\end{IEEEeqnarray}

As proved in~\cite{bashar2019uplink}, the problem \eqref{eq:prob2} can be formulated as a standard geometric programming problem, and can be solved by a software solver such as CVX for MATLAB~\cite{grant2020cvx,grant2008graph}.

\subsection{Max-Min Energy Efficiency Method}
To improve the EE at a given SE, \cite{ito2021} proposed the max-min EE TPC method. Similar to the max-min SE, the optimization problem of the max-min EE method can be written as:

\begin{IEEEeqnarray}{ll}
\maximize_{\mathopen{}\left\{q_k\right\}\mathclose{}}&\min_{k=1,\dots,K}\frac{\text{Bandwidth}\cdot S_k}{\Bar{P}q_k+P_{\text{U}}}\label{eq:mm_ee_prob}\\
\text{subject to }&S_k\ge S^{\mathopen{}\left(\text{r}\right)\mathclose{}},k=1,\dots,K\IEEEnonumber\\
&0\le q_k\le 1,k=1,\dots,K.\IEEEnonumber%
\end{IEEEeqnarray}
where $S^{\mathopen{}\left(\text{r}\right)\mathclose{}}$ is the required minimum (target) SE for UEs to ensure a certain quality of service. 

%\begin{IEEEeqnarray}{ll}
%\maximize_{\mathopen{}\left\{q_k\right\}\mathclose{}}&\min_{k=1,\dots,K}w_k^{\mathopen{}\left(\text{b}\right)\mathclose{}}E_k\label{eq:mm_ee_prob}\\
%\text{subject to }&S_k\ge %S_k^{\mathopen{}\left(\text{r}\right)\mathclose{}},k=1,\dots,K\IEEEnonumber\\
%&0\le q_k\le 1,k=1,\dots,K,\IEEEnonumber%
%\end{IEEEeqnarray}

To make the problem easier to handle, replace $q_k$ in the denominator with an auxiliary variable $\nu$:

\begin{IEEEeqnarray}{ll}
\maximize_{\mathopen{}\left\{q_k\right\}\mathclose{},\nu}&\min_{k=1,\dots,K}\frac{\text{Bandwidth}\cdot S_k}{\Bar{P}\nu+P_{\text{U}}}\label{eq:mm_ee_prob2}\\
\text{subject to }&S_k\ge S^{\mathopen{}\left(\text{r}\right)\mathclose{}},k=1,\dots,K\IEEEnonumber\\
&0\le q_k\le 1,k=1,\dots,K\IEEEnonumber\\
&q_k\le\nu,k=1,\dots,K\IEEEnonumber\\
&\nu^*\le\nu\le 1,\IEEEnonumber%
\end{IEEEeqnarray}
where $\nu^*$ is the slack variable and given as the maximum $q_k$ that achieves the target SE, obtained by solving the following optimization problem:

\begin{IEEEeqnarray}{ll}
\minimize_{\mathopen{}\left\{q_k\right\}\mathclose{}}&\max_{k=1,\dots,K}q_k\label{eq:qk}\\
\text{subject to }&S_k\ge S^{\mathopen{}\left(\text{r}\right)\mathclose{}},k=1,\dots,K\IEEEnonumber\\
&0\le q_k\le 1,k=1,\dots,K.\IEEEnonumber%
\end{IEEEeqnarray}
which is explained further in \cite{ito2021}. The optimization problem is hence summarized as:
\begin{enumerate}
\item Finding the optimal value of $\nu$ to maximize the minimum EE using a hill climbing algorithm.\footnote{The initial value of $\nu$ is set to $\nu^{*}$. The step size for each iteration is set to 0.1, and $\nu$ approaches 1. If the obtained minimum EE is smaller that that of the previous point, the step size will be divided by 2 and the sign will be inverted, i.e., the point will turn back with a smaller step. The iteration will end if the step size becomes smaller than $10^{-4}$.}
\item Optimizing $q_k$ to minimize the maximum transmit power when the target SE is reached.\footnote{While (\ref{eq:qk}) is a non-convex problem, the constraint $S^{(r)}-S_k \leq 0$ can be transformed into a polynomial function, and geometric programming can find the global optimal solution for (\ref{eq:qk}).}
\end{enumerate}

When solving the EE-maximization problem, $\nu$ is always the maximum value of $q_k$ while $\nu^*$ is the maximum value of $q_k$ that achieves the required SE. Therefore, the actual EE, which is calculated by using the optimized $q_k$, becomes higher than is calculated within the optimization problem because the actual denominator of EE for those UEs also becomes smaller ($\Bar{P}q_k+P_{\text{U}}\leq\Bar{P}\nu+P_{\text{U}})$.

\section{Channel Measurement} \label{chan_meas}
\subsection{Channel Sounder} \label{sndrr}
\begin{figure}[t!]
     \centering
     \includegraphics[width=0.75\linewidth]{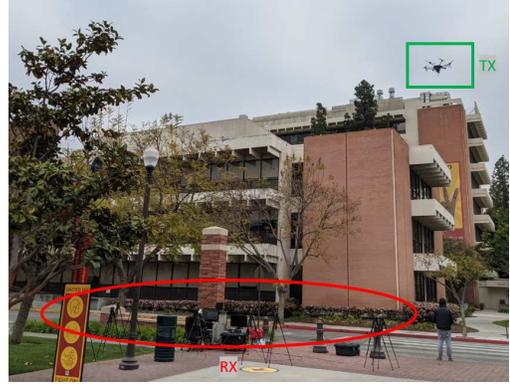}
     \caption{Channel sounder setup: a single TX antenna on a drone passes through all possible AP locations in the air while 8 separated RX antennas on camera stands are positioned at 8 different UE locations.}
     \label{fig:sndr}
\end{figure}

We acquired our channel data with a measurement setup (``channel sounder'') that includes a transmitter (TX) on a drone \cite{ponce2021air} and a receiver (RX) on the ground (Fig. \ref{fig:sndr}).\footnote{We emphasize that we are not considering a drone-based wireless system, but we are only obtaining the channel data for CF-mMIMO systems using the drone channel sounder for system analysis in Sec. \ref{perf}.} The TX, which sends out a known waveform, acts as a \emph{virtual array} with a single omnidirectional antenna being moved by the drone along a trajectory of different possible AP locations, while the RX remains stationary at one location. The RX, which records the received waveforms for later postprocessing, is connected via an RF switch to eight physically separated omnidirectional antennas,\footnote{The separations of the eight antennas per RX site are limited to 15m by the RF cables connecting the antennas to the switch.} and thus records the channels for 8 UEs during each measurement run. Note that while our setup measures the downlink channel, the resulting channel measurements can still be used for uplink performance evaluations, since propagation channels are reciprocal \cite{molisch2011wireless}. 

The signal is a 46 MHz OFDM-like sounding waveform with 2301 subcarriers (20 kHz subcarrier spacing) at 3.5 GHz.\footnote{We do not compensate for the carrier frequency and sampling rate offsets, and assume they are parts of the channels. In fact, we consider the channel coefficients at each subcarrier as a particular channel realization of a single frequency (3.5 GHz flat channel).} As the TX moves along a trajectory at 1 m/s speed, it continuously transmits the waveform at 27 dBm while the RX constantly captures the channel data between the TX antenna and 8 RX antennas every 50 ms through switching. Therefore, 1$\times$8$\times$2301 channel matrix may be captured at every 5 cm of drone movement. The characteristics of the drone sounder and the channel measurement principle for CF-mMIMO systems are further discussed in \cite{choi2021using, ponce2021air}.

The measured channel may show correlation in the fading at the different antenna elements, in particular when they are closely spaced together. However, due to the measurement principle of our sounder, phase coherence of the measurements between closely located points could not be achieved. This is partly due to the oscillator drift during the time that it takes the drone to move between locations, the potential positioning error, and the vibrations. Consequently, it is difficult to consider a correlation model for the linear array with closely spaced antenna elements. In order to resolve this issue, we have decided to separate each antennas by at least four wavelengths during evaluations, even for the cases of co-located scenarios and semi-distributed scenarios. This achieves greater diversity of the antenna arrays (which is beneficial for deployment), and ensures that the signals have uncorrelated phases at the antenna elements both in theory and in the measurements.

\begin{figure}[t!]
     \centering
     \includegraphics[width=0.75\linewidth]{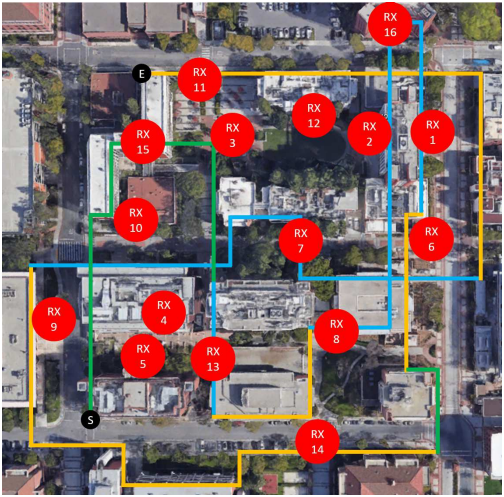}
     \caption{Overall channel measurement setting: the routes of the TX around 200m$\times$200m area at USC are shown by the yellow, green, and blue lines, which represent trajectories at 25m, 35m, and 45m heights, and the same trajectories were repeated 16 times for 16 different RX sites marked in red, resulting in channel data for $\sim$30,000 possible AP and 128 UE locations.}
     \label{env}
\end{figure}

\subsection{Channel Measurement Setting} \label{setting}
The channel measurements were conducted at the southwest side of the University of Southern California (USC) University Park Campus, in Los Angeles, CA, USA, as shown in Fig. \ref{env}. The service area, i.e., area in which the (virtual) APs and the UEs are deployed, is a $200$m$\times 200$m area. The drone passes through all locations at 25m, 35m, and 45m heights where the APs may potentially be installed, varying across the trajectory depending on the height of the building the drone flies over, autonomously by a software app. 
For each measurement, a RX site was selected within the service area to place the 8 UE antennas at 1.5m height (either in open space, under trees, under a building, or inside a building). After a round of trajectory measurements was completed at one RX site, the RX sounder was placed at a different site to repeat the same process. As described in Sec. \ref{sndrr}, the channel data is captured every 5cm and there were 8 UEs antennas per RX site. Since the drone TX sounder moved through a $\sim$1500m trajectory and there were in total 16 RX sites, channel data between $\sim$30,000 possible AP and 128 UE locations were obtained.\footnote{The channel environment may show some variations during the repeated flights along the trajectory, and the path of the drone may not exactly overlap for different RX sites due to elongated measurement time. Please refer \cite{choi2021using} for a discussion and measures taken to minimize the impact of such variations.}

\subsection{Applying Measurement Data to Analysis}
In our analysis, we can obtain 2301 different realizations of $\bm{\mathrm{H}}$ from the measurement campaign between $M$ BS antennas chosen from the trajectory and $K$ UEs chosen from 128 possible locations, since our analysis focuses on frequency-flat channels. Hence, the channel coefficients at 2301 subcarriers are regarded as particular realizations of a 3.5 GHz flat channel. If we define each frequency index as $i$, then $\beta_{m,k} = \frac{1}{F}\sum_{i=1}^{F}h_{m,k}(i)^2$ is the average large-scale fading with $F$ the total number of frequency indices ($F=2301$). Calibration and time invariance of the sounder characteristics over the duration of the measurements were tested, and some frequency points (less than 10\% of the total acquired data) exhibiting excessive calibration errors were discarded.

\begin{figure}[t!]
     \centering
     \includegraphics[width=0.9\linewidth]{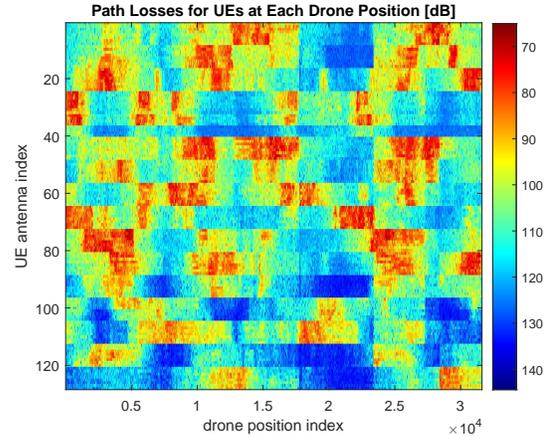}
     \caption{Path losses between the drone flying across the TX trajectory and 128 UE antennas distributed across 16 RX sites at 3.5 GHz frequency.}
     \label{fig:PL}
\end{figure}

\subsection{Path Losses From Propagation}
In order to show some propagation characteristics from the channel measurement campaigns, Fig. \ref{fig:PL} shows the path losses between the drone flying across the TX trajectory (with starting and end positions given in Fig. \ref{env}), which are about 30,000 potential AP antenna locations, and 128 UE antennas distributed across 16 RX sites.\footnote{Some discontinuities within the trajectory (along the x-axis) in Fig. \ref{fig:PL} come from removal of erroneous points.} The SNR of the measurement varied, and was as high as 50 dB: we could measure path loss between about $70$ dB to $120$ dB after compensating for the hardware calibration. It shows that the path losses for most of the UEs close to each other, either within a single RX site (in groups of 8 UE antennas) or RX sites close to one another, have similar values, while the path losses change drastically if the UE antennas are far from one another. %A closer look at the propagation characteristics such as frequency and spatial correlation is out of scope of this paper and will be presented in future work.

\subsection{Comparison With Rayleigh Model}
\begin{figure}[t!]
     \centering
     \includegraphics[width=0.95\linewidth]{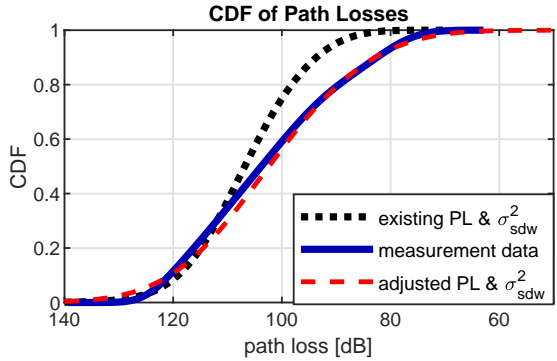}
     \caption{Comparison of a) Rayleigh model using typical path loss law and shadowing channel parameters, b) measurement data, and c) Rayleigh model with channel parameters adjusted to the measurement data.}
     \label{fig:siso}
\end{figure}

Fig. \ref{fig:siso} shows the raw measurement data compared with Rayleigh model that uses a standard $\alpha$-$\beta$ path loss law plus log-normal shadowing. Specifically, we compared with the cases when A) the path loss (in dB) is $\mathrm{L}(d) = 30.5+36.7\mathrm{log}_{10}(d)$ where $d$ is the distance between an AP and a UE (in meters) and shadowing standard deviation value is 4 dB ($\sigma_{sdw}^2 = 4~\mathrm{dB}$) as suggested from \cite{demir2020foundations} and B) an ``adjusted model'' that obtains the path loss law and shadowing obtained from direct fitting of all our measurement data: $\mathrm{L}(d) = 68.3568+52.3\mathrm{log}_{10}(d/25)$ and $\sigma^2_{sdw} = 9~\mathrm{dB}$. 25m breakpoint from the adjusted model comes from the minimum distance between the TX and the RX during the measurement.

The gap is shown not only between the measurement data and the existing model, but also in between the measurement data and the adjusted model. This comes from a fundamental model mismatch, i.e., the {\em structure} of the popular $\alpha$-$\beta$ plus log-normal shadowing model does not fit completely to the structure of the measured data. In particular, the adjusted model can achieve an unrealistically low path loss ($<60~\mathrm{dB}$) due to high variance in shadowing, which can dominate the system performance in MIMO scenarios if some of the channels between a UE and APs are channels with unusually low path loss. Development of a more detailed channel model for distributed MIMO that can explain all the observed features of path loss and correlation is a subject for future work.

\section{Performance Evaluations} \label{perf}
Performances of the TPC algorithms in Sec. \ref{TPC} are evaluated and compared by applying them to the channel data obtained from the measurement campaigns described in Sec. \ref{chan_meas}, using various setup parameters. We assume MMSE combining unless MR is mentioned specifically. We fix 20 MHz of bandwidth, 290K noise temperature, and 7 dB noise figure for all simulations. For the transmit power ($\bar{P}$) and the circuit power ($P_\text{U}$), 0.2W \cite{dahlman20185g} and 0.1W \cite{hao2018energy} are assumed. For the max-min EE TPC algorithm, we consider low target SE to maximize the EE unless stated otherwise.

\begin{figure}[!t]
    \centering
    \subfloat[CDF of spectral efficiency]{\includegraphics[width=0.95\linewidth, keepaspectratio]{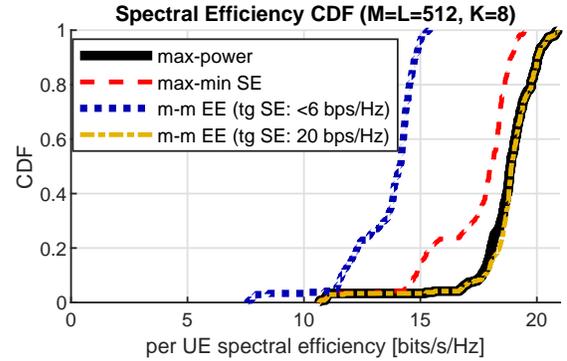}
    \label{fig:SE}}%\\[-2ex]
    \\
    \subfloat[CDF of energy efficiency]{\includegraphics[width=0.95\linewidth, keepaspectratio]{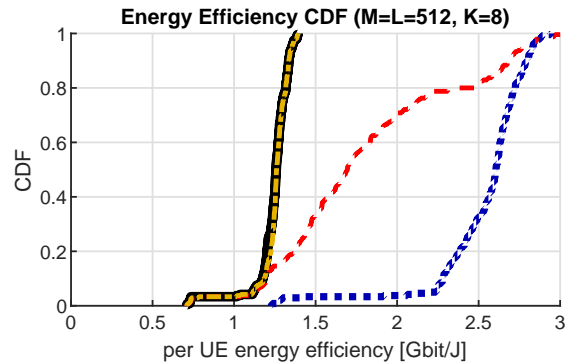}
    \label{fig:EE}}
    
    \caption{CDFs of spectral and energy efficiency for three different TPC algorithms, when 512 single-antenna APs serve 8 UEs.}
    \label{fig:SE_EE}
\end{figure}

\subsection{Comparing Different Energy Efficiency Algorithms}
First, we compare different types of TPC algorithms and evaluate their trade-offs. For this comparison, fully-distributed 512 single-antenna APs ($M=L=512$) and 8 single-antenna UEs ($K=8$) are chosen randomly from 30,000 and 128 possible locations respectively. 

Cumulative distribution functions (CDFs) in Fig. \ref{fig:SE_EE} show that the SE generally ranks in the order of max-power, max-min SE, and max-min EE algorithm (with target SE$<$6 bits/s/Hz), while the EE ranks conversely, so that there is a clear trade-off between the SE and EE. It must be noted that the performance of the max-min EE algorithm can differ significantly depending on the target SE parameter. For example, if we compare the max-min EE plots with two different target SE values (20 and $<$6)\footnote{Results for the max-min EE are the same if the target SE less than 6 bits/s/Hz; the EE does not increase further with decreasing SE, so decreasing the target SE further will not change the result.} on Fig. \ref{fig:SE_EE}, the median of the SE is greater for 20 bits/s/Hz target SE by about 4.8 bits/s/Hz, while the median of the the EE is 1.3 Gbit/J less than that of $<$6 bits/s/Hz target SE. Both the SE and EE plots for the max-min EE approach with 20 bits/s/Hz target SE overlap on the plots of the max-power algorithm.\footnote{This is limited to MMSE combining, as will be shown later.} 

The max-min EE hence is a very flexible algorithm where the target SE acts as an adjustable parameter modifying the system performances depending on the SE or EE requirements. However, the disadvantage of the max-min EE algorithm is its runtime, requiring high computing power to be used in real time.\footnote{If the run time of computing the UE transmit power for the max-min SE algorithm is scaled to 1, the max-min EE algorithm in average takes about 16 times longer than the max-min SE algorithm.} The max-min SE meanwhile provides the middle ground performance between the max-min EE with target SE $<$6 bits/s/Hz and max-power.

\subsection{Impacts of Serving Indoor UEs From Outdoor APs} \label{indr}
It is noteworthy that for the max-min EE and max-min SE algorithms, a small step-like behavior in the CDF occurs near the 20-30\% level. This occurs because there is a 23\% probability that the randomly chosen set of UEs involves at least one indoor UE (associated with 4 antennas located indoors at RX5 site on Fig. \ref{env}, also corresponding to UE37 to UE40 in Fig. \ref{fig:PL}), which has a poor channel quality, impacting the TPC algorithm for all other UEs in the same set (remember that the TPC algorithm maximizes the {\em minimum} EE performance). Likewise, for the max-power, there are about 3 percent UEs with very low SE, which comes from the $4/128$ probability of selecting an indoor UE. 

\begin{figure}[!t]
    \centering
    \subfloat[CDF of spectral efficiency]{\includegraphics[width=0.95\linewidth, keepaspectratio]{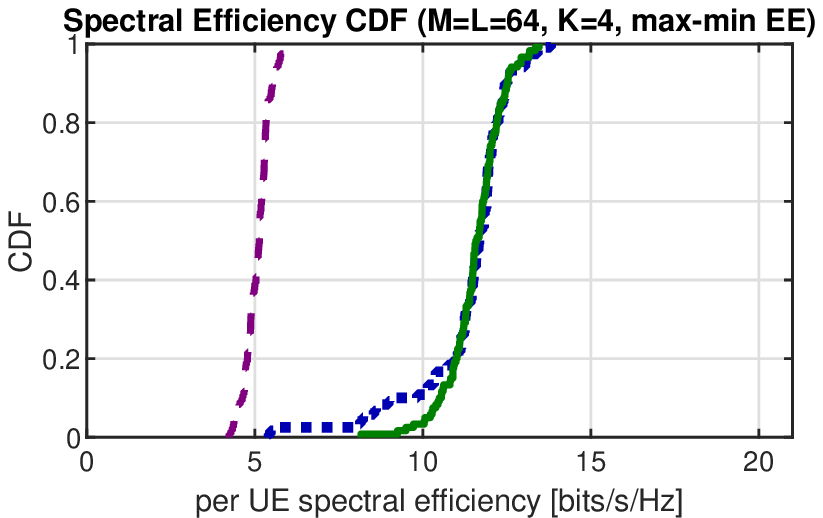}
    \label{fig:SE2}}%\\[-2ex]
    \\
    \subfloat[CDF of energy efficiency]{\includegraphics[width=0.95\linewidth, keepaspectratio]{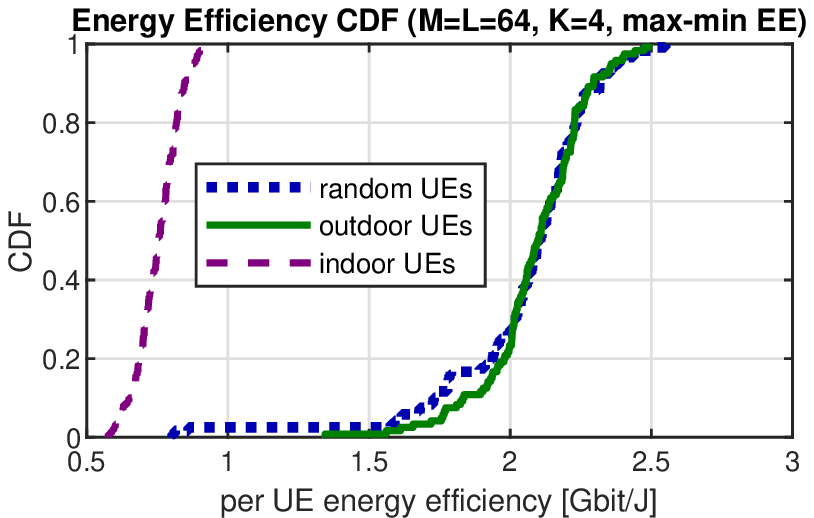}
    \label{fig:EE2}}
    
    \caption{CDFs of spectral and energy efficiency for max-min EE, when 64 single-antenna APs serve 4 UEs selected from different environments.}
    \label{fig:SE_EE2}
\end{figure}

Fig. \ref{fig:SE_EE2} shows the case with 64 single-antenna APs and 4 UEs, where the max-min EE algorithm with $<$6 bits/s/Hz target SE is applied. While the horizontal tail and the stepping behavior at the lower end of the CDF in the case when we randomly select UEs are not as evident as in the case of $K=8$ from Fig. \ref{fig:SE_EE} (due to 12\% of selecting at least one indoor UE from the measurement data), such characteristics still remain. These characteristics are removed and the CDF is smoother if we only select outdoor UEs. In contrast, if only the 4 indoor UEs are served, the performances for both the SE and EE are very poor. This shows that the performance of TPC algorithms can be heavily affected by the UEs with poor channel qualities.

\begin{figure*}[!t]
    \centering
    \subfloat[spectral efficiency (max-power)]{\includegraphics[width=0.45\linewidth]{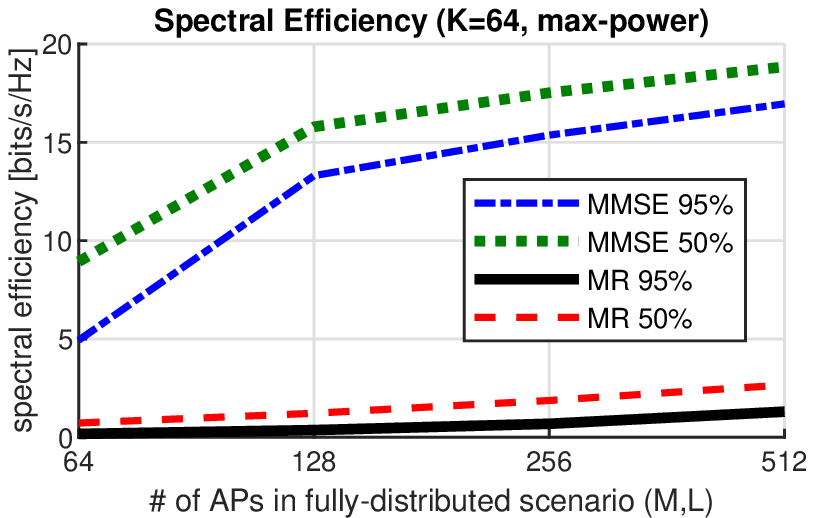}\hfill%
    \label{fig:SE_M}}
    \subfloat[energy efficiency (max-power)]{\includegraphics[width=0.45\linewidth]{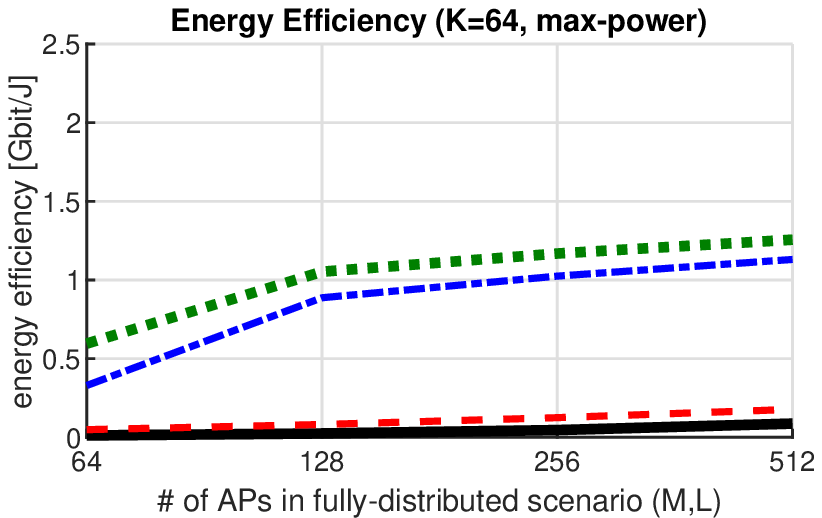}\label{fig:EE_M}}%\\[-2ex]
    \\
    \subfloat[spectral efficiency (max-min EE)]{\includegraphics[width=0.45\linewidth]{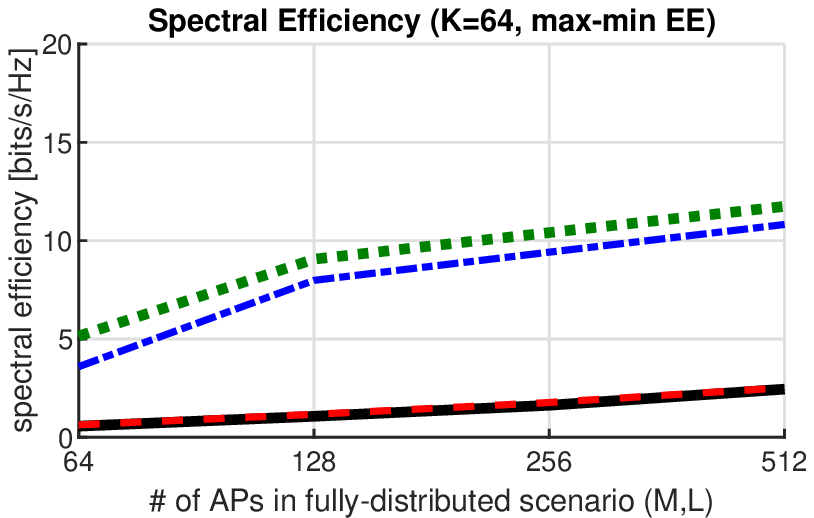}%
    \label{fig:SE_M-m}}
    \subfloat[energy efficiency (max-min EE)]{\includegraphics[width=0.45\linewidth]{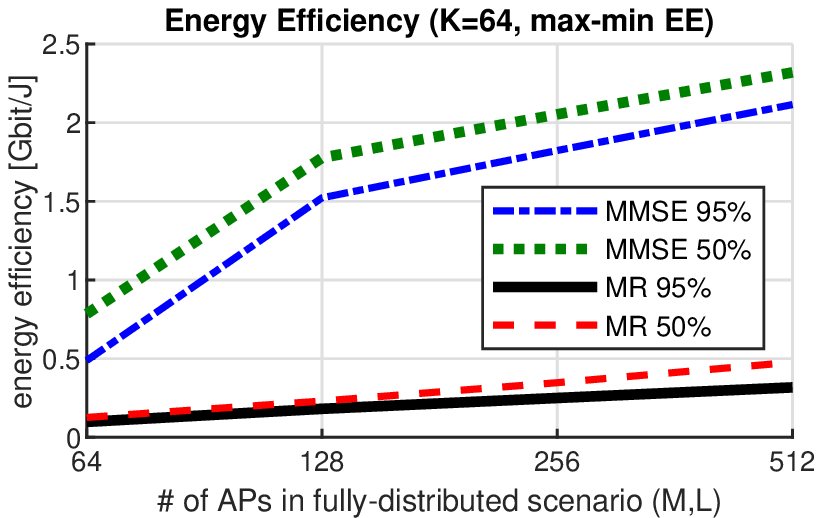}%
    \label{fig:EE_M-m}}    
    \caption{Spectral and energy efficiency of the fully-distributed scenario when the number of UEs is fixed at 64 ($K=64$) and the number of BS antennas (APs) varies from 64 to 512 ($M=L=64,128,256,512$) - two TPC algorithms (max-power and max-min EE) are compared.}
    \label{fig:SE_EE_M}
\end{figure*}

\begin{figure*}[!t]
    \centering
    \subfloat[spectral efficiency (max-power)]{\includegraphics[width=0.45\linewidth]{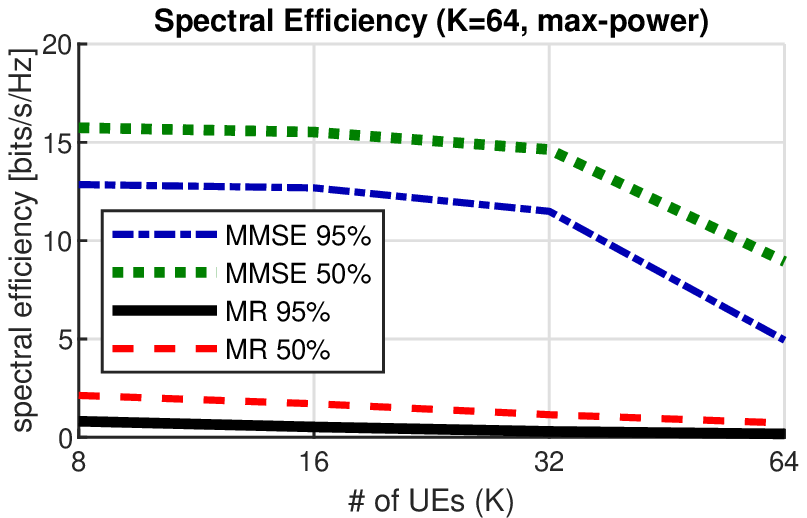}%
    \label{fig:SE_K}}
    \subfloat[energy efficiency (max-power)]{\includegraphics[width=0.45\linewidth]{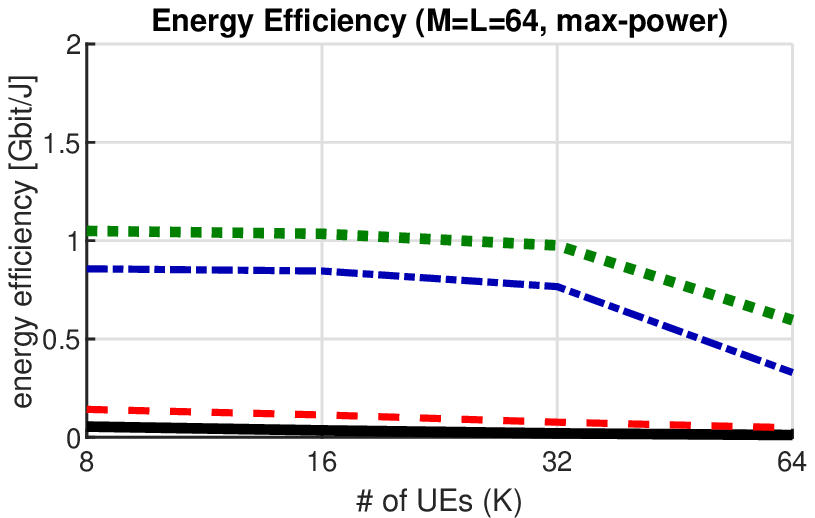}\label{fig:EE_K}}
    \\
    \centering
    \subfloat[spectral efficiency (max-min EE)]{\includegraphics[width=0.45\linewidth]{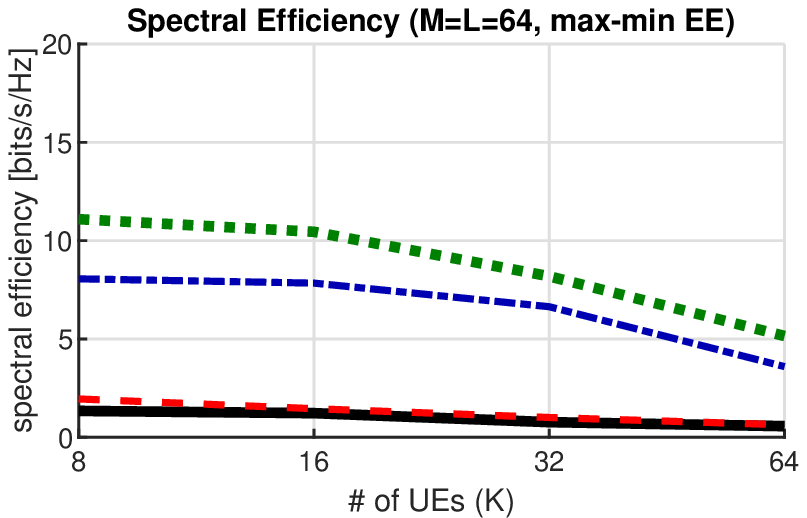}%
    \label{fig:SE_K-m}}
    \subfloat[energy efficiency (max-min EE)]{\includegraphics[width=0.45\linewidth]{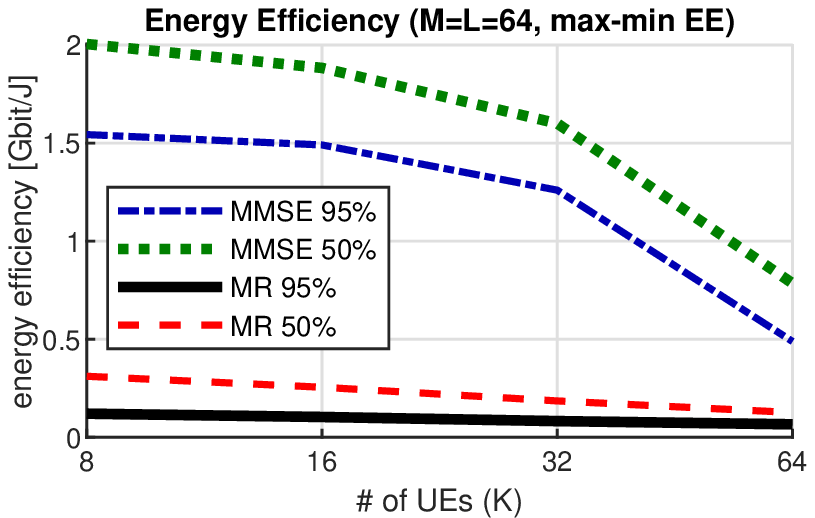}%
    \label{fig:EE_K-m}}    
    \caption{Spectral and energy efficiency of the fully-distributed scenario when the number of BS antenna is fixed at 64 ($M=L=64$) and the number of UEs varies from 8 to 64 ($K=8,16,32,64$) - two TPC algorithms (max-power and max-min EE) are compared.}
    \label{fig:SE_EE_K}
\end{figure*}

\subsection{MMSE vs. MR: Varying the Number of BS Antennas}
Fig. \ref{fig:SE_EE_M} shows results for a larger number of UEs ($K=64$), which creates a more challenging scenario. We still assume the fully-distributed scenario, and the number of BS antennas (single-antenna APs) varies from 64 to 512 ($M=L=64,128,256,512)$, but consider only the max-power and max-min EE algorithms. Both the MMSE and MR are compared, by looking at the median (50\% likely) and lower-end (95\% likely) values of the CDFs per scenario.

In the max-power case on Fig. \ref{fig:SE_M} and Fig. \ref{fig:EE_M}, both the SE and EE increase with the number of BS antennas, for both MMSE and MR. The increase is the largest when moving from $M=64$ to $M=128$ for the MMSE because the MMSE, although much better than MR, is not as effective when $M=K$, since this usually leads to an ill-conditioned channel matrix and therefore excessive noise enhancement \cite{molisch2011wireless}.
%There is also a significant gap between the 95\% curve and the 50\% curve {\em for all numbers of BS antennas}. This is in contrast to a concentrated array, where the inherent diversity of a larger array should lead to a reduced gap between the curves. 
Comparing the MR against the MMSE, the MR has much worse performance due to its inability to cancel interference. Performance is increased slightly as $M$ increases, but remains far below the MMSE performance. This indicates that even extremely large arrays ($M=512$) do not provide the theoretically predicted similar performance between the MMSE and MR combining. 

The results for the max-min EE are shown on Fig. \ref{fig:SE_M-m} and Fig. \ref{fig:EE_M-m}. %First of all, the 95\% likely value and 50\% likely values are similar, indicating more uniform performance than for max-power transmission.
Compared to the max-power, the EE increases much faster with the number of BS antennas when MMSE is used. For MR combining, in contrast, the max-min EE algorithm has close or better performance to the max-power algorithm for both the SE and EE. This is because the MR performance is dominated by interference, and the max-min EE algorithm controls the power coefficients of the UEs, which can result in both the interference mitigation and the energy reduction. However, the performance of MR is still much worse than the MMSE. In terms of the \emph{total system} EE, this indicates a trade-off in the energy consumption of the UE, and the energy consumption at the receiving BS, since MMSE requires more energy both for the more complicated processing, and the backhauling of the received data to a central processing location. However, the details of this optimization will depend significantly on the specific processing, backhauling hardware, and the relative importance the network operator assigns to UE and infrastructure energy consumption. 

\begin{figure*}[!t]
    \centering
    \subfloat[CDF of spectral efficiency (max-power)]{\includegraphics[width=0.45\linewidth]{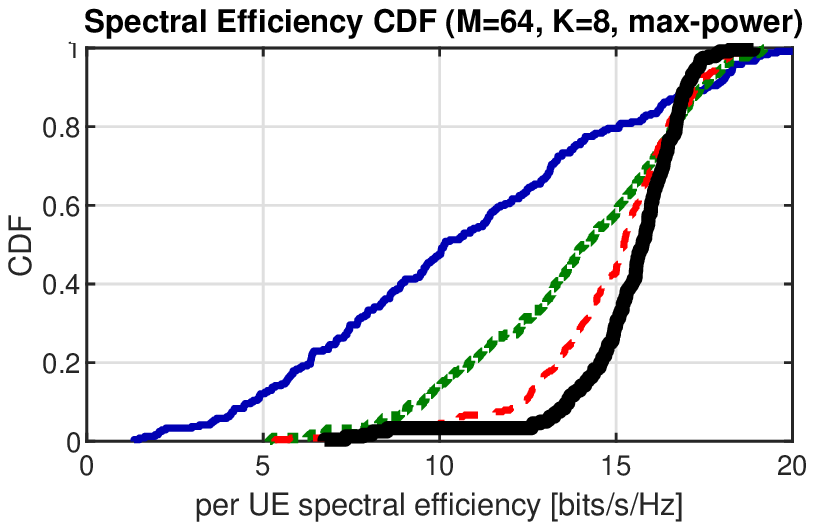}%
    \label{fig:SE_S}}
    \subfloat[CDF of energy efficiency (max-power)]{\includegraphics[width=0.45\linewidth]{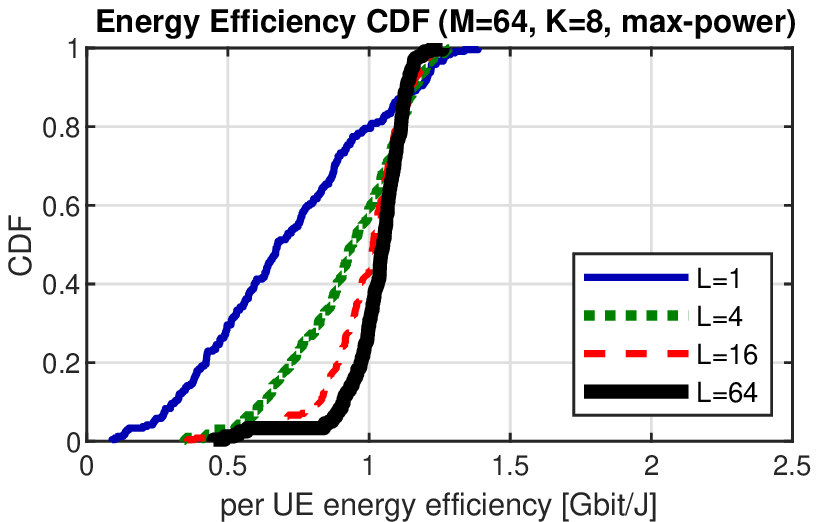}\label{fig:EE_S}}
    \\
    \centering
    \subfloat[CDF of spectral efficiency (max-min EE)]{\includegraphics[width=0.45\linewidth]{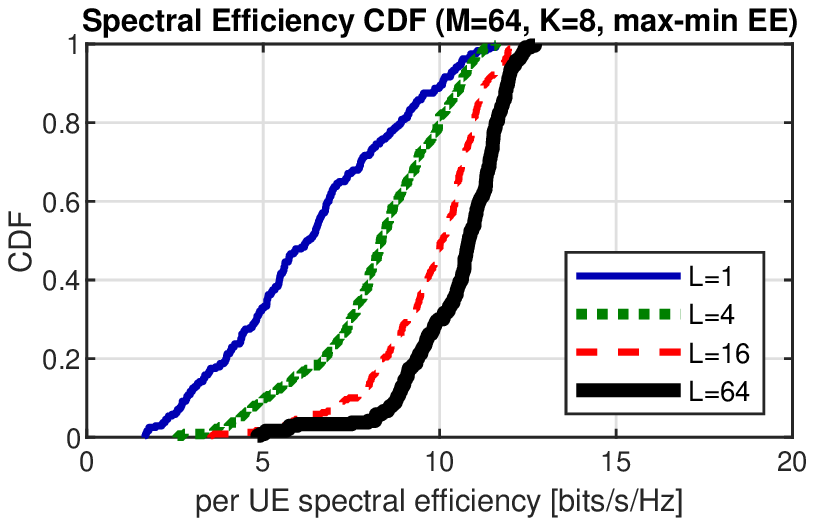}%
    \label{fig:SE_S-m}}
    \subfloat[CDF of energy efficiency (max-min EE)]{\includegraphics[width=0.45\linewidth]{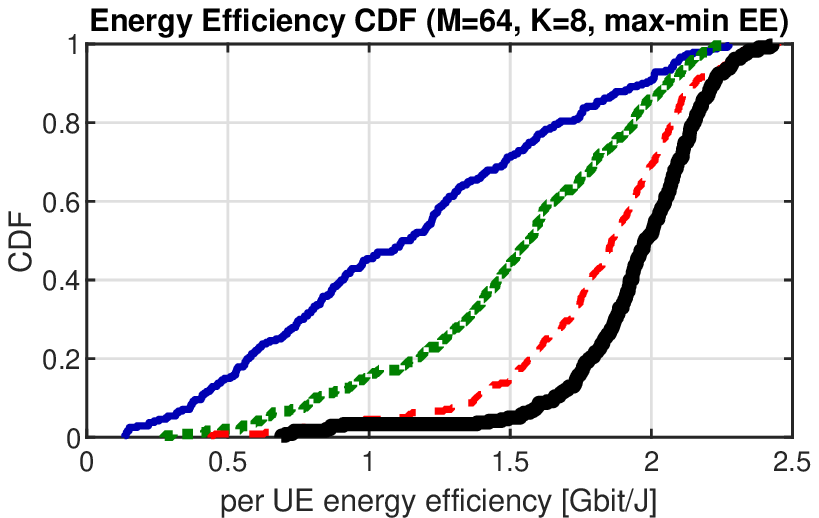}%
    \label{fig:EE_S-m}}    
    \caption{CDF of spectral and energy efficiency where the number of APs varies from 1 to 64 ($L=1,4,16,64$) while the total number of BS antennas and UEs remain the same ($K=8$) - two TPC algorithms (max-power and max-min EE) are compared.}
    \label{fig:SE_EE_s}
\end{figure*}

\subsection{MMSE vs. MR: Varying the Number of UEs}
We compare how both the SE and EE changes if the number of BS antennas is fixed and the number of UEs increases. We consider only the fully-distributed case with $M=L=64$ and $K=8,16,32,64$. Again, for the TPC algorithms, the max-power and max-min EE are used, and both MMSE and MR processing are analyzed. The medians (50\% likely) and lower-ends (90\% likely) of the CDFs are shown in Fig. \ref{fig:SE_EE_K}.

We first look at the max-power algorithm. Looking at MMSE, the performance decreases slowly with increasing K, except for the case when $M=K=64$, where the decrease is sharp due to difficulty of cancelling interference. %does not change much even when the number of UEs increases, because MMSE can effectively cancel out interference. In contrast, 
Meanwhile, the performance also decreases for MR, as the interference from other UEs increases with the number of UEs. We also see that the SE performance gap between MMSE and MR remains very large even when we have many more APs than UEs: $M=64$ and $K=8$. 
%Comparing the MMSE and MR, the performance gap is large even when the number of base station antennas is large and the number of UEs is small in contrast to the conception that the MR performs as well as MMSE with large number of base station antennas. 

Now we look at the max-min EE algorithm. Compared with the max-power method, the EE increased at the cost of SE for the MMSE, but both the SE and EE values are equal or greater for the MR. This again is because each UE does not transmit at the full power, so the reduction in interference helps the SE while also improving the EE. Another difference to the max-power is that the 50\% likely performance of the MMSE decreases more sharply with the number of UEs, even when $M>K$. In summary, the performances of both max-power and max-min EE decrease with the number of UEs, the MMSE is much more effective than MR despite its complexity, and the max-min EE algorithm is again shown to be effective in improving the EE, especially when the number of UEs is much less than the number of BS antennas.

\subsection{Comparing Different Number of APs}
We fix the total number of BS antennas ($M=64$) and vary the number of APs ($L$) to find the best way to deploy the APs among fully-distributed ($L=M$), semi-distributed ($1<L<M$), and co-located ($L=1$) cases, similar to the evaluations for indoor scenarios in \cite{choi2020co-located}. The number of UEs is fixed at 8 ($K=8$). We compare the max-power and max-min EE TPC algorithm when $L=1,4,16,64$.\footnote{For the semi-distributed and co-located cases, we select consecutive spatial points of the drone which are separated by at least 43cm, which is about 5$\lambda$ at 3.5 GHz carrier frequency.}

The results in Fig. \ref{fig:SE_EE_s} show that for the max-power algorithm, fully-distributed ($L=64$) deployment has the best SE and EE for about 80\% of the UEs, while the co-located case ($L=1$) has the best peak performance. This contrasts with our previous results from \cite{choi2020co-located}, where the semi-distributed performance was very close to fully-distributed in a smaller indoor environment with fewer BS antennas ($M=8$). The CDFs are also steeper when there are more APs ($L$). This makes sense because distributing more APs across areas allows UEs to have at least one good channel between all BS antennas and a UE. In contrast, a UE is likely to experience a bad channel with all BS antennas in the co-located case when the single AP is strongly shadowed from the UE; yet this case can also result in the highest performance if it has good channels to all antennas of the AP.

Even for the max-min EE, the fully-distributed case still performs the best. All cases resulted in better EE in comparison to max-power at the cost of SE, and more performance gap in EE could be made with more APs ($L$). In summary, more APs usually provides the better SE and EE, and the max-min EE is especially more useful when there are more APs. 

\begin{figure*}[!t]
    \centering
    \subfloat[CDF of spectral efficiency (max-power)]{\includegraphics[width=0.45\linewidth]{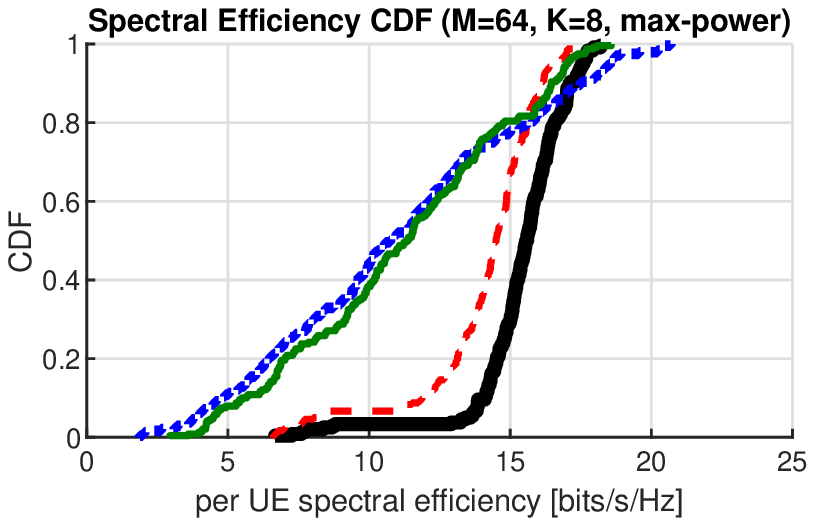}%
    \label{fig:SE_C}}
    \subfloat[CDF of energy efficiency (max-power)]{\includegraphics[width=0.45\linewidth]{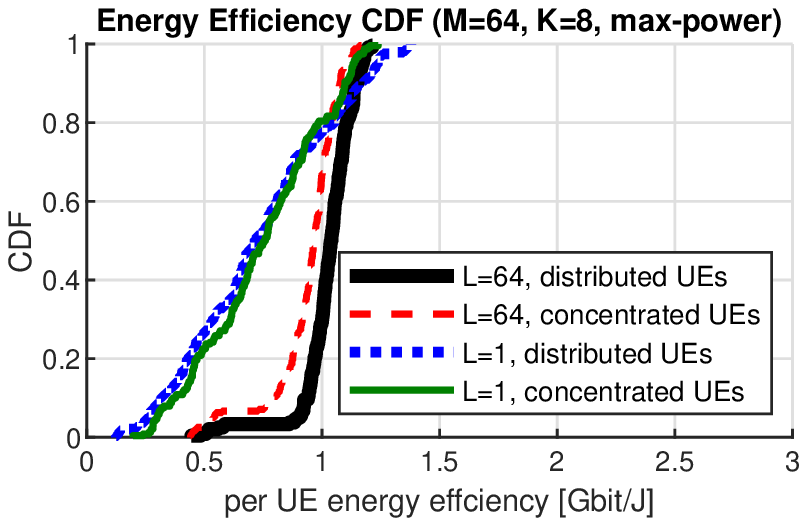}\label{fig:EE_C}}
    \\
    \centering
    \subfloat[CDF of spectral efficiency (max-min EE)]{\includegraphics[width=0.45\linewidth]{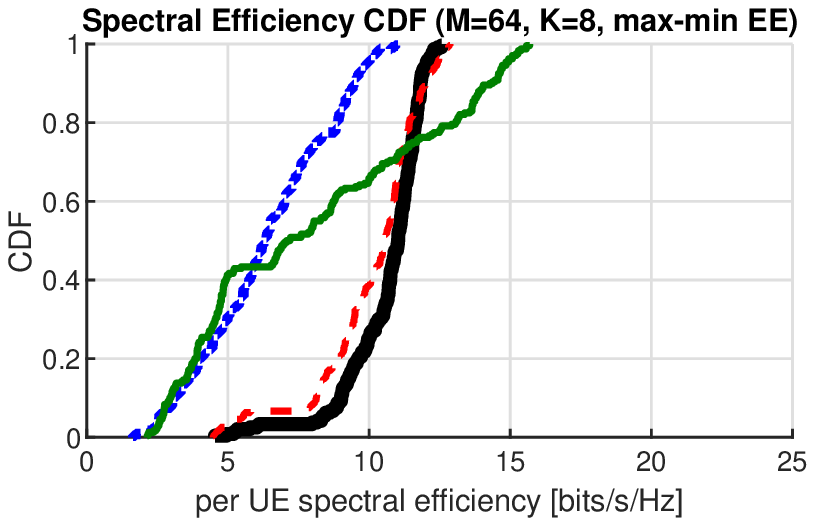}%
    \label{fig:SE_C-m}}
    \subfloat[CDF of energy efficiency (max-min EE)]{\includegraphics[width=0.45\linewidth]{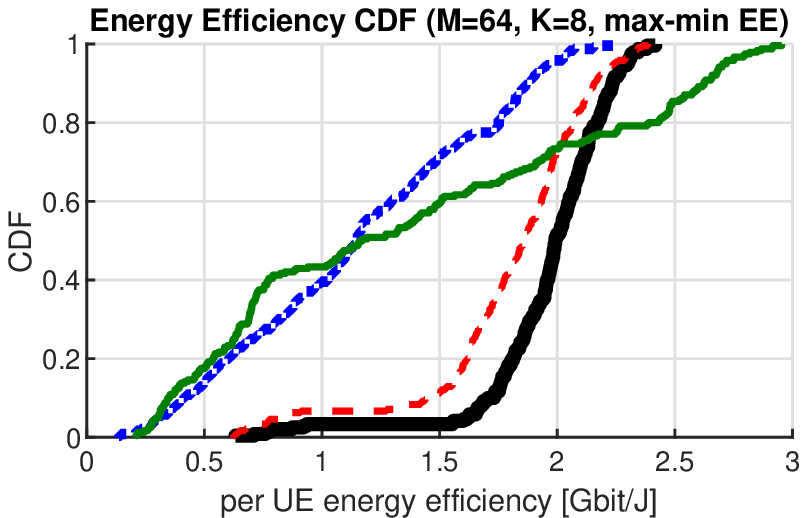}%
    \label{fig:EE_C-m}} 
    \caption{CDFs of spectral and energy efficiency for co-located ($L=1$), and fully-distributed ($L=64$) scenarios, when the UEs are either concentrated to one site versus when the UEs are distributed to different sites - two TPC algorithms (max-power and max-min EE) are compared.}
    \label{fig:SE_EE_C}
\end{figure*}

\begin{figure*}[!t]
    \centering
    \subfloat[CDF of spectral efficiency (max-power)]{\includegraphics[width=0.45\linewidth]{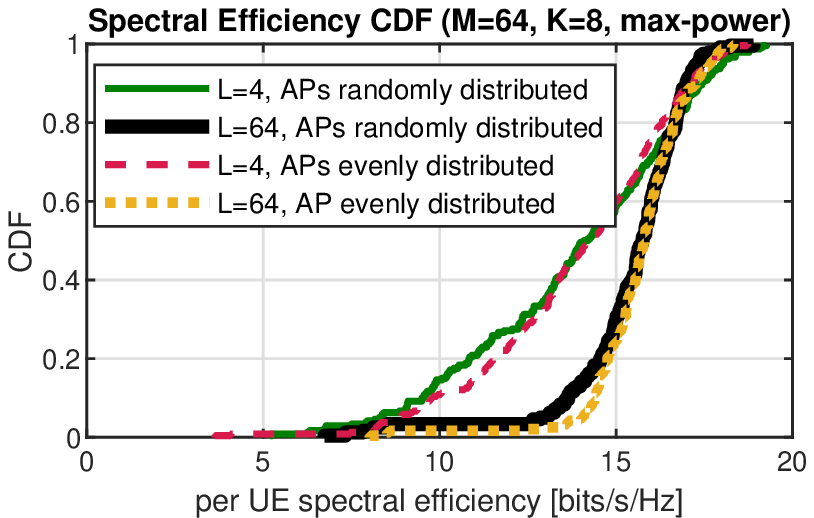}%
    \label{fig:SE_E}}
    \subfloat[CDF of energy efficiency (max-power)]{\includegraphics[width=0.45\linewidth]{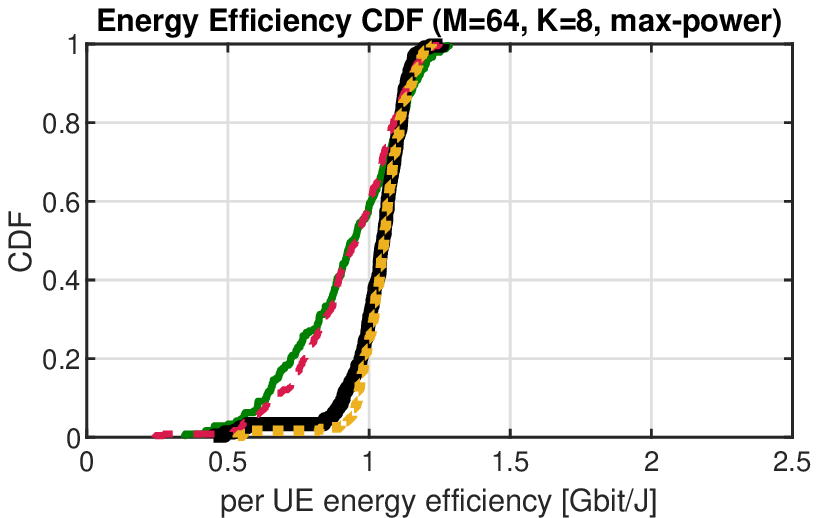}\label{fig:EE_E}}
    \\
    \centering
    \subfloat[CDF of spectral efficiency (max-min EE)]{\includegraphics[width=0.45\linewidth]{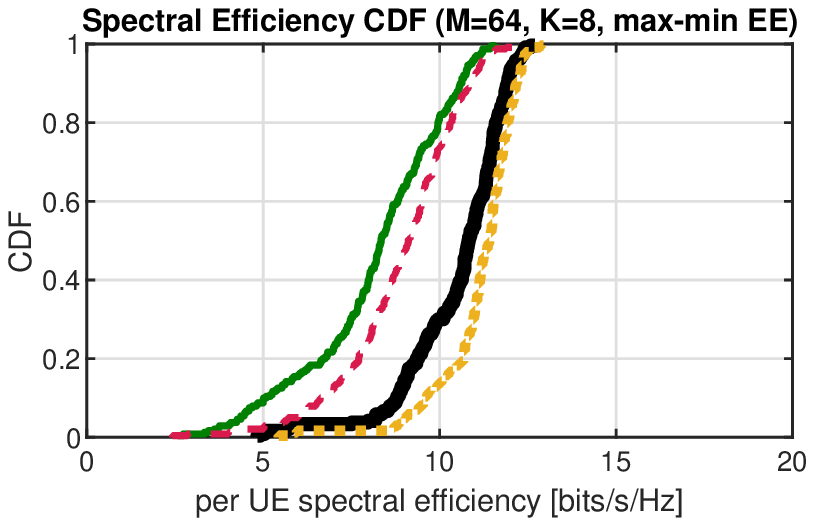}%
    \label{fig:SE_E-m}}
    \subfloat[CDF of energy efficiency (max-min EE)]{\includegraphics[width=0.45\linewidth]{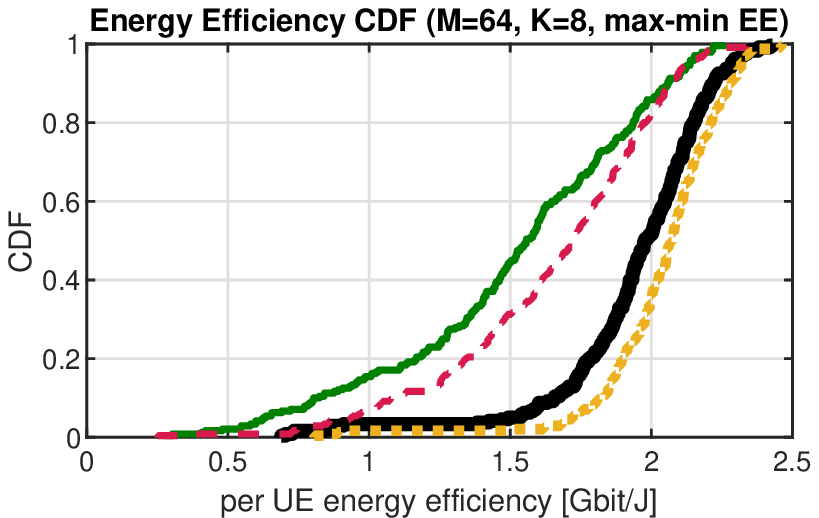}%
    \label{fig:EE_E-m}} 
    \caption{CDFs of spectral and energy efficiency when for semi-distributed ($L=4$) and fully-distributed ($L=64$) scenarios, when the APs are located either randomly versus evenly across the coverage area - two TPC algorithms (max-power and max-min EE) are compared.}
    \label{fig:SE_EE_e}
\end{figure*}

\subsection{Comparing Different UE Concentrations}
In order to determine how the performance differs depending on possible clustering of the UEs, we compare the cases when 8 UEs ($K=8$) are randomly distributed versus when 8 UEs are concentrated to a single RX site from Sec. \ref{setting} (at most 15m distance between the UEs). We fix the number of BS antennas to 64, and the BS antennas are either co-located ($L=1$) or fully-distributed ($L=64$). We again compare max-power and max-min EE TPC algorithms by observing the CDFs for SE and EE shown on Fig. \ref{fig:SE_EE_C}.

For the max-power case, we notice that the performance is better when the UEs are distributed across different sites when the BS antennas are fully-distributed because the geometric separation translates to an easier separation in the angle domain. At the same time, different sets of BS antenna are dominant for different UEs, creating an almost-block-diagonal structure of the $\bm{\mathrm{H}}$ matrix that is advantageous for the condition number and thus reduces noise enhancement. Meanwhile, the co-located scenario shows that the performances are similar for both the distributed UEs and concentrated UEs, as it has both the advantage of ability to separate the UEs as well as the disadvantage of higher chance of shadowing.

In contrast, the max-min EE algorithm interestingly shows that for the co-located scenario, further increase in the EE can be achieved when the UEs are concentrated than when UEs are distributed. The uniformity of the channels works in favor for co-located BS when assigning the transmit power coefficients for the max-min EE. For the fully-distributed scenario, concentrated UEs still perform little worse than the distributed UEs.

\subsection{Comparing Different AP Locations}
Until now, all the APs were selected at a random from all possible locations. We compare such random placement of the APs to the regular placement of the APs, i.e., dividing up the coverage area to $L$ different grid areas, and selecting an AP per grid area. 

Fig. \ref{fig:SE_EE_e} shows the case where $M=64$ and the number of APs varies ($L=4,64$) when there are 8 UEs ($K=8$). The max-power case shows that the evenly spaced APs provide similar performance as the randomly spaced APs. In contrast, the max-min EE case shows that the evenly distributed APs generally provides a larger increase in performance than randomly distributed APs. Overall, spacing the APs evenly per area is recommended, especially if the max-min EE algorithm is used, but strict planning of the deployment may not be necessary.

\section{Conclusion}
For CF-mMIMO, evaluating the trade-offs between the SE and EE for different types of TPC algorithms is very important for a large number of battery-powered UEs which the system serves. Our work shows that the max-min EE algorithm can be very effective in comparison to the max-power or max-min SE algorithm in terms of improving EE, based on the channel data obtained from extensive measurement campaigns. The analysis showed that the algorithm is more effective when no UE within a set of served UEs is in a bad channel condition, MMSE combining is applied, when the number of UEs are much less than the number of BS antennas, when the BS antennas are fully-distributed with even spacing, and when the UEs are distributed in the case of distributed BS antennas. Overall, the max-min EE is expected to improve the EE for future CF-mMIMO systems, when very high SEs from the UEs are not required.

\bibliography{IEEEabrv,reference.bib}

% Generated by IEEEtran.bst, version: 1.14 (2015/08/26)
\begin{thebibliography}{10}
\providecommand{\url}[1]{#1}
\csname url@samestyle\endcsname
\providecommand{\newblock}{\relax}
\providecommand{\bibinfo}[2]{#2}
\providecommand{\BIBentrySTDinterwordspacing}{\spaceskip=0pt\relax}
\providecommand{\BIBentryALTinterwordstretchfactor}{4}
\providecommand{\BIBentryALTinterwordspacing}{\spaceskip=\fontdimen2\font plus
\BIBentryALTinterwordstretchfactor\fontdimen3\font minus
  \fontdimen4\font\relax}
\providecommand{\BIBforeignlanguage}[2]{{%
\expandafter\ifx\csname l@#1\endcsname\relax
\typeout{** WARNING: IEEEtran.bst: No hyphenation pattern has been}%
\typeout{** loaded for the language `#1'. Using the pattern for}%
\typeout{** the default language instead.}%
\else
\language=\csname l@#1\endcsname
\fi
#2}}
\providecommand{\BIBdecl}{\relax}
\BIBdecl

\bibitem{demir2020foundations}
{\"O}.~T. Demir, E.~Björnson, and L.~Sanguinetti, ``Foundations of
  user-centric cell-free massive {MIMO},'' \emph{Found. Trends® Signal
  Process.}, vol.~14, no. 3-4, Jan. 2021.

\bibitem{choi2021using}
T.~Choi \emph{et~al.}, ``Using a drone sounder to measure channels for
  cell-free massive {MIMO} systems,'' \emph{arXiv preprint arXiv:2106.15276},
  2021.

\bibitem{zhao2015analysis}
L.~Zhao, K.~Li, K.~Zheng, and M.~Omair~Ahmad, ``An analysis of the tradeoff
  between the energy and spectrum efficiencies in an uplink massive {MIMO-OFDM}
  system,'' \emph{{IEEE} Trans. Circuits Syst. {II}}, vol.~62, no.~3, pp.
  291--295, 2015.

\bibitem{saatlou2018power}
O.~Saatlou, M.~O. Ahmad, and M.~N.~S. Swamy, ``Power control for massive
  multiuser {MIMO} systems with finite-dimensional channel,'' in \emph{2018
  16th IEEE Int. New Circuits Syst. Conf. (NEWCAS)}, pp. 18--21.

\bibitem{saatlou2019joint}
------, ``Joint data and pilot power allocation for massive {MU-MIMO} downlink
  {TDD} systems,'' \emph{{IEEE} Trans. Circuits Syst. {II}}, vol.~66, no.~3,
  pp. 512--516, 2019.

\bibitem{hao2018energy}
H.~Li, J.~Guo, Y.~Wang, L.~Li, Z.~Wang, H.~Wang, and J.~Gao, ``Energy efficient
  antenna selection scheme for downlink massive {MIMO} systems,'' in \emph{2018
  IEEE Int. Symp. Circuits Syst. (ISCAS)}, pp. 1--4.

\bibitem{nguyen2017energy}
L.~D. Nguyen, T.~Q. Duong, H.~Q. Ngo, and K.~Tourki, ``Energy efficiency in
  cell-free massive {MIMO} with zero-forcing precoding design,'' \emph{{IEEE}
  Commun. Lett.}, vol.~21, no.~8, pp. 1871--1874, 2017.

\bibitem{ngo2018on}
H.~Q. Ngo, L.-N. Tran, T.~Q. Duong, M.~Matthaiou, and E.~G. Larsson, ``On the
  total energy efficiency of cell-free massive {MIMO},'' \emph{{IEEE} Trans.
  Green Commun. Netw.}, vol.~2, no.~1, pp. 25--39, 2018.

\bibitem{alonzo2018energy}
M.~Alonzo, S.~Buzzi, and A.~Zappone, ``Energy-efficient downlink power control
  in {mmWave} cell-free and user-centric massive {MIMO},'' in \emph{2018 IEEE
  5G World Forum (5GWF)}, pp. 493--496.

\bibitem{tran2019first}
L.-N. Tran and H.~Q. Ngo, ``First-order methods for energy-efficient power
  control in cell-free massive {MIMO}: invited paper,'' in \emph{2019 53rd
  Asilomar Conf. Signals Syst. Comput.}, pp. 848--852.

\bibitem{jin2020spectral}
S.-N. Jin, D.-W. Yue, and H.~H. Nguyen, ``Spectral and energy efficiency in
  cell-free massive {MIMO} systems over correlated {Rician} fading,''
  \emph{IEEE Syst. J.}, pp. 1--12, 2020.

\bibitem{qiu2020downlink}
J.~Qiu, K.~Xu, X.~Xia, Z.~Shen, and W.~Xie, ``Downlink power optimization for
  cell-free massive {MIMO} over spatially correlated {Rayleigh} fading
  channels,'' \emph{IEEE Access}, vol.~8, pp. 56\,214--56\,227, 2020.

\bibitem{zhang2018performance}
J.~Zhang, Y.~Wei, E.~Björnson, Y.~Han, and S.~Jin, ``Performance analysis and
  power control of cell-free massive {MIMO} systems with hardware
  impairments,'' \emph{IEEE Access}, vol.~6, pp. 55\,302--55\,314, 2018.

\bibitem{yang2018energy}
H.~Yang and T.~L. Marzetta, ``Energy efficiency of massive {MIMO}: cell-free
  vs. cellular,'' in \emph{2018 IEEE 87th Veh. Technol. Conf. (VTC Spring)},
  pp. 1--5.

\bibitem{alageli2020optimal}
M.~Alageli, A.~Ikhlef, F.~Alsifiany, M.~A.~M. Abdullah, G.~Chen, and
  J.~Chambers, ``Optimal downlink transmission for cell-free {SWIPT} massive
  {MIMO} systems with active eavesdropping,'' \emph{{IEEE} Trans. Inf.
  Forensics Security}, vol.~15, pp. 1983--1998, 2020.

\bibitem{bai2019max-min}
X.~Bai, Y.~Zhang, M.~Zhou, X.~Qiao, and L.~Yang, ``Max-min fairness for
  cell-free massive {MIMO} with zero-forcing receiver,'' in \emph{2019 IEEE 5th
  Int. Conf. Comp. Commun. (ICCC)}, pp. 2034--2038.

\bibitem{zhao2020efficient}
X.~Zhao, J.~Zhang, J.~Zhang, F.~Xiong, and B.~Ai, ``Efficient receiver for
  cell-free massive {MIMO} systems with low-resolution {ADCs},'' in \emph{2020
  IEEE Int. Conf. Commun. Workshops (ICC Workshops)}, pp. 1--6.

\bibitem{zhang2020analysis}
Y.~Zhang, Y.~Cheng, M.~Zhou, L.~Yang, and H.~Zhu, ``Analysis of uplink
  cell-free massive {MIMO} system with mixed-{ADC/DAC} receiver,'' \emph{IEEE
  Syst. J.}, pp. 1--12, 2020.

\bibitem{demir2021joint}
{\"O}.~T. Demir and E.~Björnson, ``Joint power control and {LSFD} for
  wireless-powered cell-free massive {MIMO},'' \emph{{IEEE} Trans. Wireless
  Commun.}, vol.~20, no.~3, pp. 1756--1769, 2021.

\bibitem{yan2021scalable}
H.~Yan, A.~Ashikhmin, and H.~Yang, ``A scalable and energy efficient {IoT}
  system supported by cell-free massive {MIMO},'' \emph{IEEE Internet Things
  J.}, pp. 1--1, 2021.

\bibitem{nguyen2020on}
H.~V. Nguyen \emph{et~al.}, ``On the spectral and energy efficiencies of
  full-duplex cell-free massive {MIMO},'' \emph{{IEEE} J. Sel. Areas Commun.},
  vol.~38, no.~8, pp. 1698--1718, 2020.

\bibitem{wang2020wirelessly}
X.~Wang, A.~Ashikhmin, and X.~Wang, ``Wirelessly powered cell-free {IoT}:
  analysis and optimization,'' \emph{IEEE Internet Things J.}, vol.~7, no.~9,
  pp. 8384--8396, 2020.

\bibitem{zhang2021spectral}
X.~Zhang, H.~Qi, X.~Zhang, and L.~Han, ``Spectral efficiency improvement and
  power control optimization of massive {MIMO} networks,'' \emph{IEEE Access},
  vol.~9, pp. 11\,523--11\,532, 2021.

\bibitem{ito2021}
M.~Ito \emph{et~al.}, ``Evaluation on energy efficiency of {UE} in {UL}
  cell-free massive {MIMO} system with power control methods,'' in \emph{2021
  IEEE GLOBECOM Workshops}.

\bibitem{wang2021live}
D.~Wang, C.~Zhang, Z.~Ji, Y.~Du, J.~Zhao, M.~Jiang, and X.~You, ``Live
  demonstration: a cloud-based cell-free distributed massive {MIMO} system,''
  in \emph{2021 IEEE Int. Symp. Circuits Syst. (ISCAS)}, pp. 1--1.

\bibitem{choi2021uplink}
T.~Choi \emph{et~al.}, ``Uplink energy efficiency of cell-free massive {MIMO}
  with transmit power control in measured propagation channels,'' in \emph{2020
  IEEE Workshop Signal Process. Syst. (SiPS)}.

\bibitem{ohurley2020comparison}
S.~O'Hurley and L.-N. Tran, ``A comparison of the uplink performance of
  cell-free massive {MIMO} using three linear combining schemes: full-pilot
  zero forcing with access point selection, matched-filter and
  local-minimum-mean-square error,'' in \emph{2020 31st Irish Signals Syst.
  Conf. (ISSC)}, pp. 1--6.

\bibitem{marzetta2010noncooperative}
T.~L. Marzetta, ``Noncooperative cellular wireless with unlimited numbers of
  base station antennas,'' \emph{{IEEE} Trans. Wireless Commun.}, vol.~9,
  no.~11, pp. 3590--3600, 2010.

\bibitem{shepard2018argos}
C.~Shepard, R.~Doost-Mohammady, J.~Ding, R.~E. Guerra, and L.~Zhong,
  ``Argosnet: a multi-cell many-antenna {MU-MIMO} platform,'' in \emph{2018
  52nd Asilomar Conf. Signals Syst. Comput.}, pp. 2237--2241.

\bibitem{bashar2019energy}
M.~Bashar, K.~Cumanan, A.~G. Burr, H.~Q. Ngo, E.~G. Larsson, and P.~Xiao,
  ``Energy efficiency of the cell-free massive {MIMO} uplink with optimal
  uniform quantization,'' \emph{{IEEE} Trans. Green Commun. Netw.}, vol.~3,
  no.~4, pp. 971--987, 2019.

\bibitem{bashar2019uplink}
M.~Bashar, K.~Cumanan, A.~G. Burr, M.~Debbah, and H.~Q. Ngo, ``On the uplink
  max–min {SINR} of cell-free massive {MIMO} systems,'' \emph{{IEEE} Trans.
  Wireless Commun.}, vol.~18, no.~4, pp. 2021--2036, 2019.

\bibitem{grant2020cvx}
M.~Grant and S.~Boyd, ``{CVX}: Matlab software for disciplined convex
  programming, version 2.2,'' \url{http://cvxr.com/cvx}, Jan. 2020.

\bibitem{grant2008graph}
------, ``Graph implementations for nonsmooth convex programs,'' in
  \emph{Recent Advances in Learning and Control}, ser. Lecture Notes in Control
  and Information Sciences, V.~Blondel, S.~Boyd, and H.~Kimura, Eds.\hskip 1em
  plus 0.5em minus 0.4em\relax Springer-Verlag Limited, 2008, pp. 95--110,
  \url{http://stanford.edu/~boyd/graph_dcp.html}.

\bibitem{ponce2021air}
J.~G. Ponce \emph{et~al.}, ``Air-to-ground directional channel sounder with
  64-antenna dual-polarized cylindrical array,'' \emph{arXiv preprint
  arXiv:2103.09135}, 2021.

\bibitem{molisch2011wireless}
A.~F. Molisch, \emph{Wireless Communications}, 2nd~ed.\hskip 1em plus 0.5em
  minus 0.4em\relax Wiley Publishing, 2011.

\bibitem{dahlman20185g}
E.~Dahlman, S.~Parkvall, and J.~Skold, \emph{5G NR: The Next Generation
  Wireless Access Technology}, 1st~ed.\hskip 1em plus 0.5em minus 0.4em\relax
  USA: Academic Press, Inc., 2018.

\bibitem{choi2020co-located}
T.~Choi, P.~Luo, A.~Ramesh, and A.~F. Molisch, ``Co-located vs distributed vs
  semi-distributed {MIMO}: measurement-based evaluation,'' in \emph{2020 54th
  Asilomar Conf. Signals Syst. Comput.}, pp. 836--841.

\end{thebibliography}

\bibliographystyle{IEEEtran}

\begin{IEEEbiography}
    [{\includegraphics[width=1in,height=1.25in,clip,keepaspectratio]{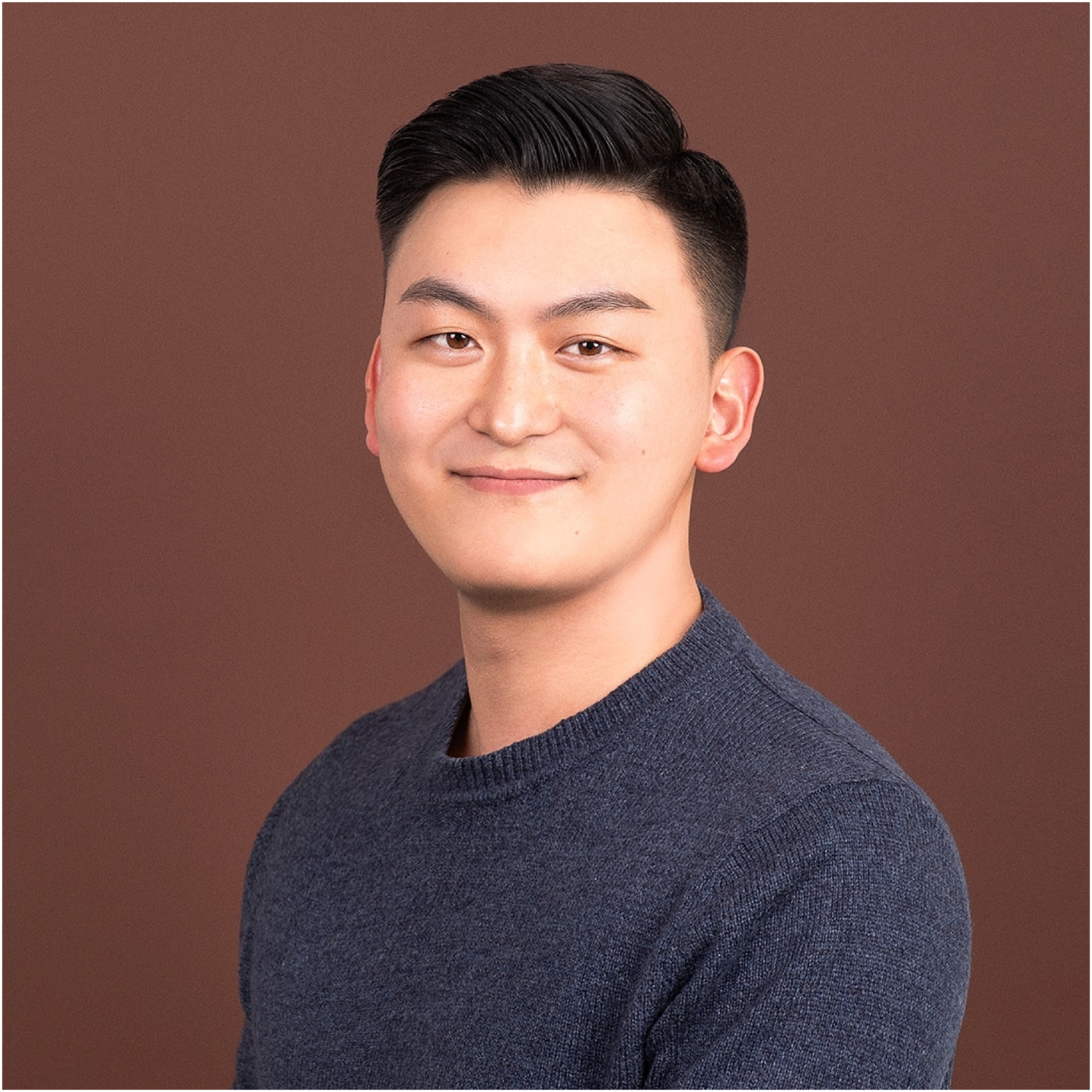}}]{Thomas Choi} (S'17) received B.S. (2015) and M.S. (2020) in Electrical Engineering from the University of Southern California (USC). He also received M.S. (2017) in Aerospace Engineering from the Georgia Institute of Technology. He is currently pursuing Ph.D. in Electrical and Computer Engineering at USC. His main research interest is verifying performances of massive MIMO technologies based on actual measured channel data, especially in the areas of distributed MIMO systems, UAV applications, and channel extrapolation.
\end{IEEEbiography}

\begin{IEEEbiography}
    [{\includegraphics[width=1in,height=1.25in,clip,keepaspectratio]{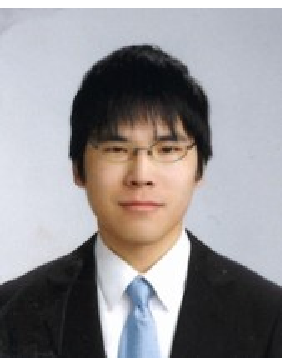}}]{Masaaki Ito} received the B.S. and M.S. degrees in wireless communications from Waseda University, Tokyo, Japan, in 2016 and 2018, respectively. From 2018 to 2019, he was a staff at KDDI Corporation, Tokyo, Japan, and was engaged in mobile network operation. He is currently an associate research engineer of wireless communications system group at KDDI Research, Inc., Saitama, Japan. His research has been concerned with user-centric networks.
\end{IEEEbiography}

\begin{IEEEbiography}
    [{\includegraphics[width=1in,height=1.25in,clip,keepaspectratio]{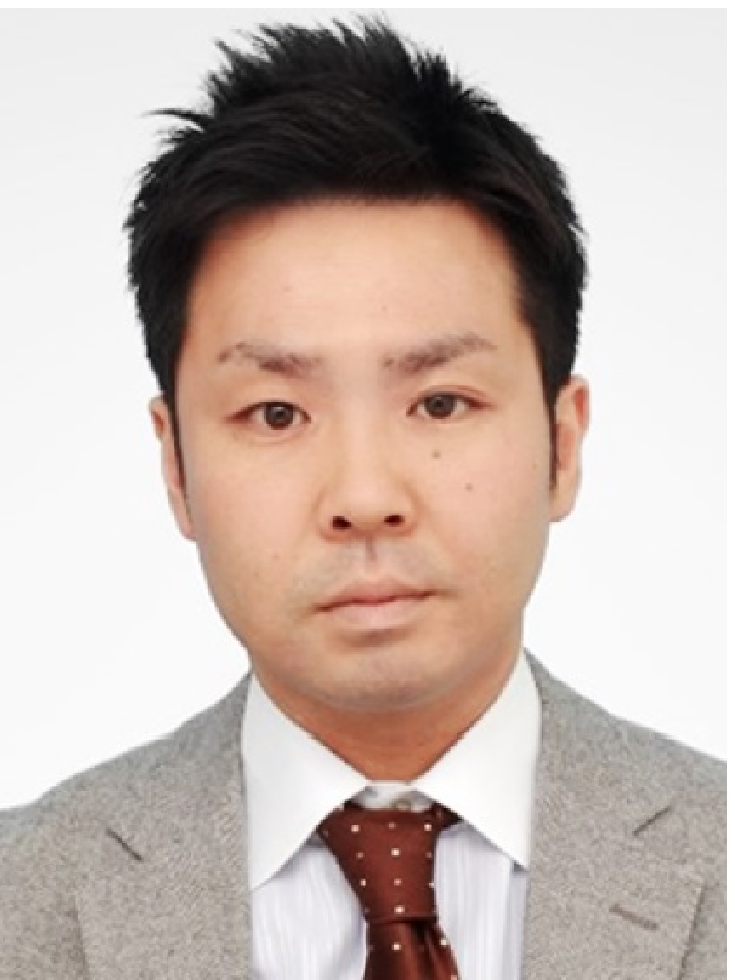}}]{Issei Kanno} received the Ph.D. degree from Tokyo Institute of Technology, Tokyo, Japan, in 2008. He then joined KDDI Corporation, where he has been engaged in research on software defned radio, antennas, and propagations in mobile communication systems. From 2013 to 2015, he was engaged in research on cognitive radio at the Advanced Telecommunication Research Institute International (ATR). Since 2015, he has been engaged in research on wireless communication systems including 5G and beyond at KDDI Research Inc. %His current research interests include signal processing and resource utilization for wireless communication systems. Dr. Kanno received the Best Paper Award at IEEE WCNC 2010 from IEEE, the Young Researchers' Award, and the Distinguished Service Award from the IEICE in 2012.
\end{IEEEbiography}

\begin{IEEEbiography}
    [{\includegraphics[width=1in,height=1.25in,clip,keepaspectratio]{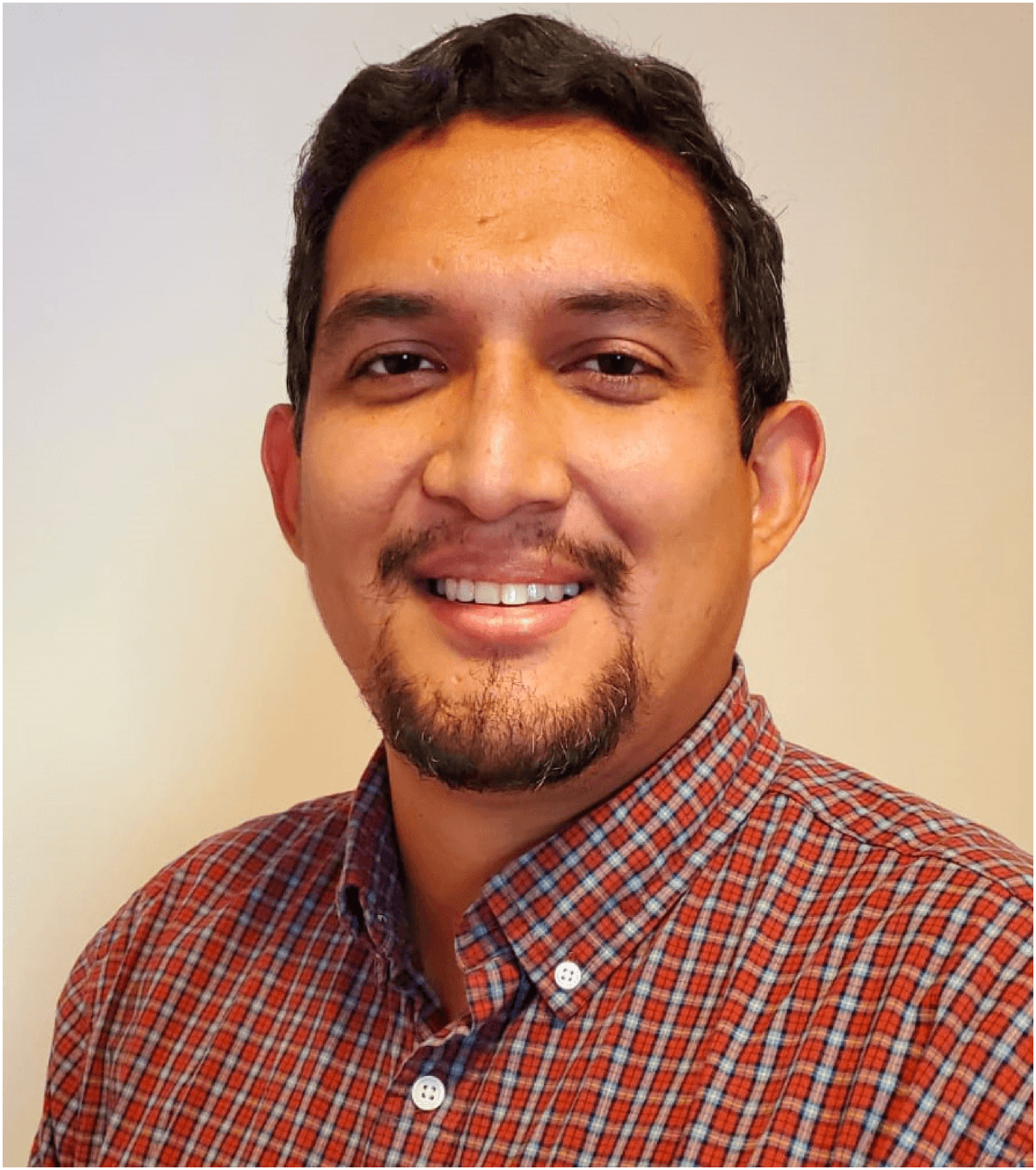}}]{Jorge Gomez Ponce} (S'17) received his B.Sc. degree in electronics and telecommunications engineering a master's degree in telecommunications from the Escuela Superior Politécnica del Litoral (ESPOL) Guayaquil, Ecuador, in 2011 and 2015, respectively. He obtained an MSEE degree in electrical engineering from the University of Southern California, Los Angeles, CA, USA, in 2019, where he is currently pursuing a Ph.D. degree in electrical engineering. %His current research interests include statistical and ray-tracer modeling for wireless communication channels (e.g., WI-Fi, UAV, mm-wave, and THz), RF systems design and implementation with DSP/FPGA systems, estimation theory and machine learning applied to wireless communications, and cognitive radios.
\end{IEEEbiography}

\begin{IEEEbiography}
    [{\includegraphics[width=1in,height=1.25in,clip,keepaspectratio]{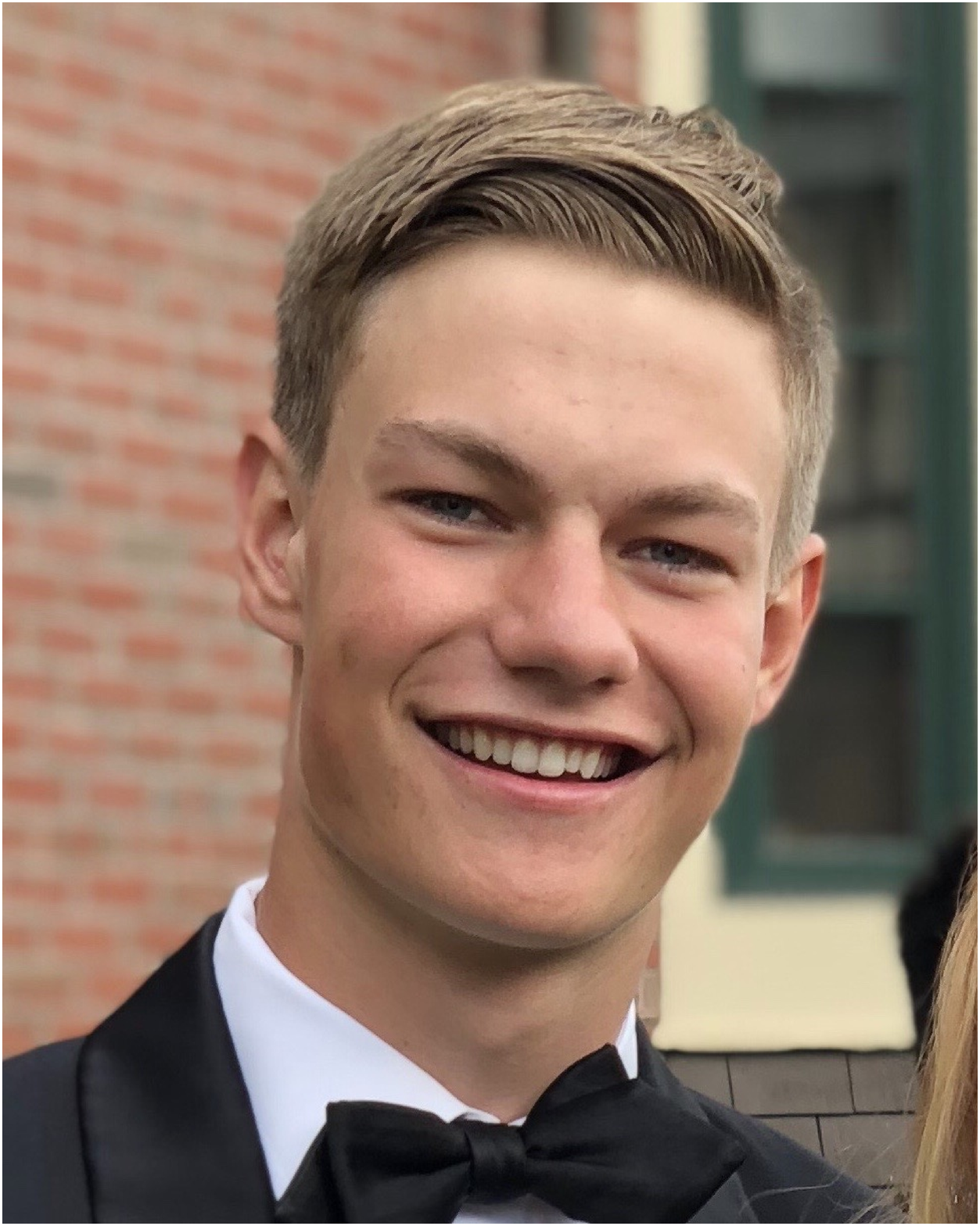}}]{Colton Bullard} will receive his B.S. degree in Aerospace Engineering in the spring of 2022. He plans to continue his education and pursue a M.S. Degree in Aerospace and Mechanical Engineering. He has previously worked in the General Aviation industry, and as an undergraduate research assistant for the USC WiDeS research laboratory. His current interests involve General Aviation propulsion systems, as well as payload systems.
\end{IEEEbiography}

\begin{IEEEbiography}
    [{\includegraphics[width=1in,height=1.25in,clip,keepaspectratio]{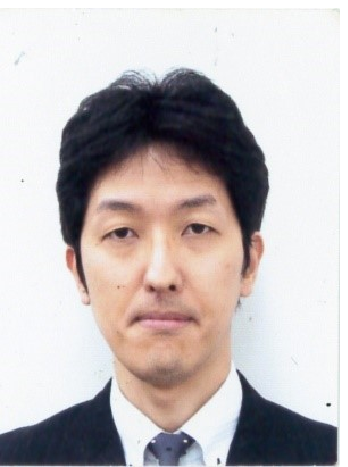}}]{Takeo Ohseki} received B.E. and M.E. degrees in electrical and electronics engineering from the Tokyo Institute of Technology, Tokyo, Japan, in 1997 and 1999 respectively. In 1999, he joined KDDI Corp. and has been engaged in research and development on wireless access and mobile communications systems. He has also been involved in the standardization work in 3GPP RAN working group 1 for many years, and received ITU-AJ Encouragement Awards in 2021. He is currently a R\&D manager at KDDI R\&D Laboratories, Inc.
\end{IEEEbiography}

\begin{IEEEbiography}
    [{\includegraphics[width=1in,height=1.25in,clip,keepaspectratio]{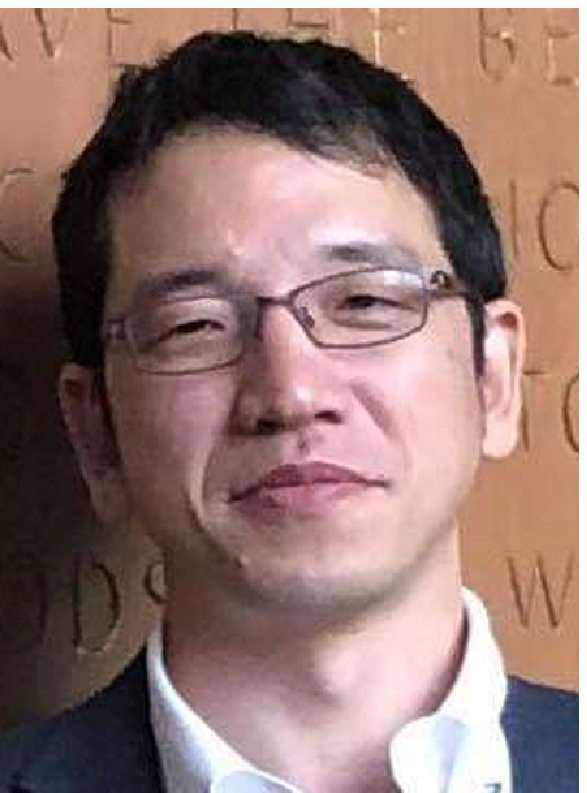}}]{Kosuke Yamazaki} received the Ph.D degree from Tokyo University, Japan in 2005. He joined KDDI and engaged in the research and development on Software Defined Radio, Cognitive Radio, WiMAX, Wi-Fi and heterogeneous networks. From 2012 to 2013, he engaged in the Consumer Marketing Division on the consumer business strategy. Since 2019, we has been the head of Wireless Communication System Laboratory on the future wireless communications system toward beyond 5G/6G.
\end{IEEEbiography}

\begin{IEEEbiography}
    [{\includegraphics[width=1in,height=1.25in,clip,keepaspectratio]{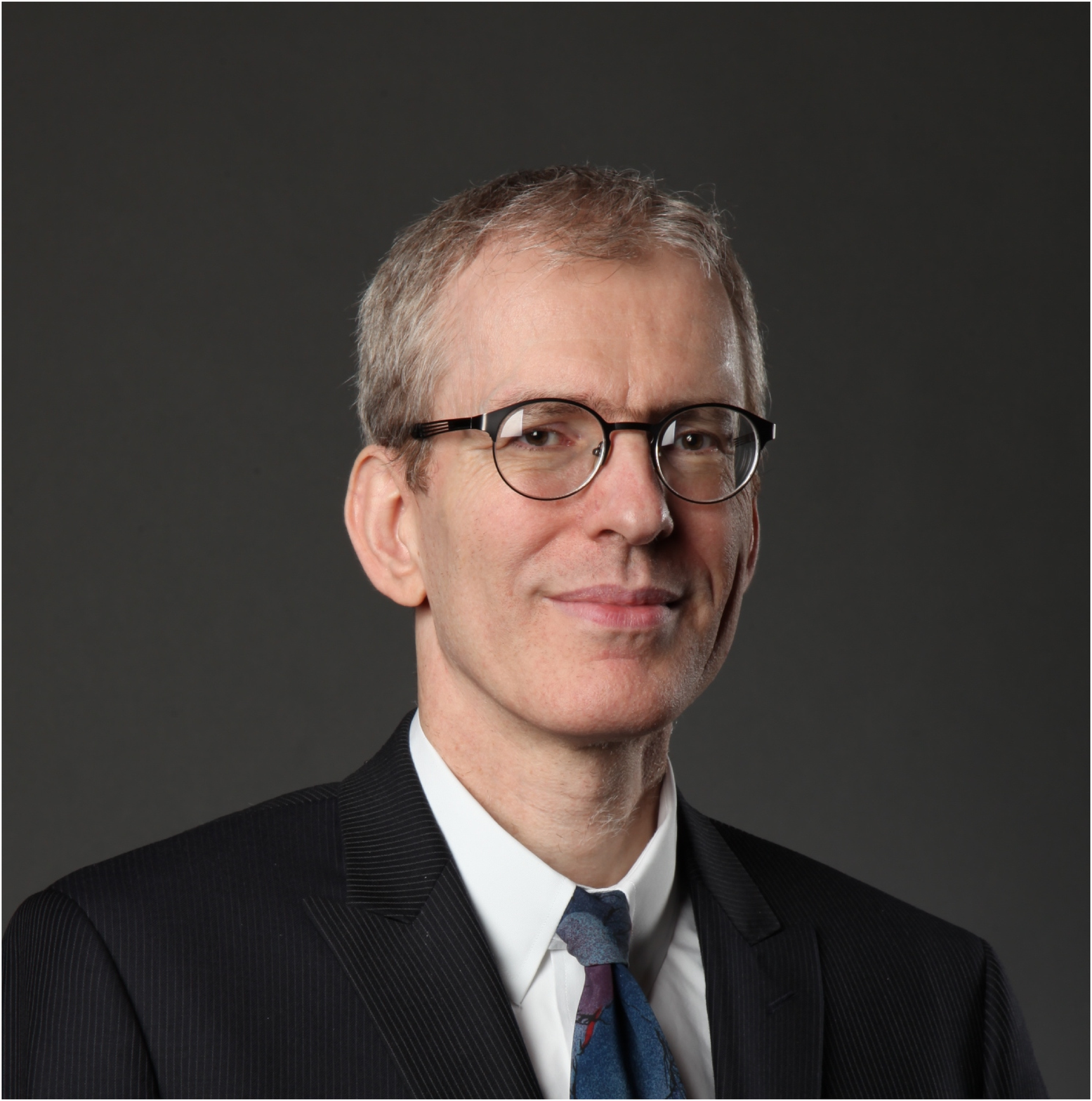}}]{Andy Molisch} received his degrees (Dipl.Ing. 1990, PhD 1994, Habilitation 1999) from the Technical University Vienna, Austria. He spent the next 10 years in industry, at FTW, AT\&T (Bell) Laboratories, and Mitsubishi Electric Research Labs (where he rose to Chief Wireless Standards Architect). In 2009 he joined the University of Southern California (USC) in Los Angeles, CA, as Professor, and founded the Wireless Devices and Systems (WiDeS) group. In 2017, he was appointed to the Solomon Golomb – Andrew and Erna Viterbi Chair. 
\end{IEEEbiography}

\end{document}